\documentclass[%
  reprint,
superscriptaddress,
nofootinbib, % show footnotes as footnotes
 amsmath,amssymb,
 aps,
]{revtex4-2}
\usepackage{graphicx}% Include figure files
\usepackage{dcolumn}% Align table columns on decimal point
\usepackage{bm}% bold math
\usepackage{multirow}
\usepackage{xspace}
\usepackage{paralist}
\usepackage{booktabs}% nice tables with toprule, bottomrule, midrule
\usepackage[hidelinks]{hyperref} % remove ugly boxes areound hyperrefs on arXiv

\newcommand{\mjj}{\ensuremath{m_\text{jj}}\xspace}
\newcommand{\pt}{\ensuremath{p_\text{T}}\xspace}
\newcommand{\kt}{\ensuremath{k_\text{t}}\xspace}
\newcommand{\ys}{\ensuremath{y^*}\xspace}
\newcommand{\yb}{\ensuremath{y_\text{b}}\xspace}
\newcommand{\ym}{\ensuremath{y_\text{max}}\xspace}
\newcommand{\yam}{\ensuremath{\left|y\right|_\text{max}}\xspace}
\newcommand{\mur}{\ensuremath{\mu_\text{R}}\xspace}
\newcommand{\muf}{\ensuremath{\mu_\text{F}}\xspace}
\newcommand{\Qsq}{\ensuremath{Q^2}\xspace}
\newcommand{\ptavg}{\ensuremath{\langle{}p_\text{T}\rangle_{1,2}}\xspace}

\newcommand{\GeV}{\ensuremath{\,\text{GeV}}\xspace}
\newcommand{\TeV}{\ensuremath{\,\text{TeV}}\xspace}
\newcommand{\chisq}{\ensuremath{\chi^{2}}\xspace}
\newcommand{\ndf}{\ensuremath{n_\text{dof}}\xspace}

\newcommand{\mz}{\ensuremath{m_\text{Z}}\xspace}
\newcommand{\as}{\ensuremath{\alpha_\text{s}}\xspace}

\newcommand{\asmz}{\ensuremath{\as(\mz)}\xspace}
\newcommand{\asmur}{\ensuremath{\as(\mur)}\xspace}

\makeatletter
\def\section{%
  \@startsection
    {section}%
    {1}%
    {\z@}% {\parindent}
    {1em \@plus1ex \@minus .2ex}%
    {-1.5em}%
    {\normalfont\normalsize\bfseries\itshape}%
}%
\def\@appendixcntformat#1{\appendixname\ \csname the#1\endcsname.~}% % insert a tilde at the end!
\def\bibsection{%
  \par
  \begingroup
  \baselineskip26\p@
  \bib@device{\hsize}{72\p@}%
  \endgroup
  \nobreak\@nobreaktrue
  \addvspace{19\p@}%
}%
\makeatother

\begin{document}

\preprint{CERN-TH-2024-224, KA-TP-25-2024, MPP-2024-260, ZU-TH/01/25}

\title{Precise Determination of the Strong Coupling Constant\\ from Dijet Cross Sections up to the Multi-TeV Range}


\author{Fazila Ahmadova}
\affiliation{ Max-Planck-Institut f\"{u}r Physik,  Boltzmannstr.\ 8, 85748 Garching, Germany\label{mpi} }
\affiliation{ Department of Physics, Universit\"{a}t Z\"{u}rich, Winterthurerstrasse 190, CH-8057 Z\"{u}rich, Switzerland  }
\author{  Daniel Britzger}
\affiliation{ Max-Planck-Institut f\"{u}r Physik,  Boltzmannstr.\ 8, 85748 Garching, Germany\label{mpi} }
\author{  Xuan Chen}
\affiliation{ School of Physics, Shandong University, Jinan, Shandong 250100, China}
\author{  Johannes Gäßler}
\affiliation{ Karlsruhe Institute of Technology (KIT), Institute for Experimental Particle Physics, Wolfgang-Gaede-Str.\ 1, 76131 Karlsruhe, Germany}
\author{  Aude~Gehrmann--De~Ridder}
\affiliation{ Institute for Theoretical Physics, ETH, Wolfgang-Pauli-Strasse 27, CH-8093 Z\"urich, Switzerland }
\affiliation{ Department of Physics, Universit\"{a}t Z\"{u}rich, Winterthurerstrasse 190, CH-8057 Z\"{u}rich, Switzerland  }
\author{  Thomas~Gehrmann}
\affiliation{ Department of Physics, Universit\"{a}t Z\"{u}rich, Winterthurerstrasse 190, CH-8057 Z\"{u}rich, Switzerland  }
\author{  Nigel~Glover}
\affiliation{ Institute for Particle Physics Phenomenology, University of Durham, Durham, DH1 3LE, United Kingdom}
\author{  Claire~Gwenlan}
\affiliation{ Department of Physics, The University of Oxford, Oxford, OX1 3PU, United Kingdom }
\author{  Gudrun~Heinrich}
\affiliation{ Karlsruhe Institute of Technology (KIT), Institute for Theoretical Physics, Wolfgang-Gaede-Str.\ 1, 76131 Karlsruhe, Germany}
\author{  Alexander~Huss}
\affiliation{ Theoretical Physics Department, CERN, CH-1211 Geneva 23, Switzerland  }
\author{  Lucas~Kunz}
\affiliation{ Karlsruhe Institute of Technology (KIT), Institute for Theoretical Physics, Wolfgang-Gaede-Str.\ 1, 76131 Karlsruhe, Germany}
\author{  Jo\~{a}o Pires}
\affiliation{  LIP, Avenida Professor Gama Pinto 2, P-1649-003 Lisboa, Portugal}
\affiliation{ Faculdade de Ciências, Universidade de Lisboa, 1749-016 Lisboa, Portugal }
\author{  Klaus Rabbertz}
\affiliation{ Karlsruhe Institute of Technology (KIT), Institute for Experimental Particle Physics, Wolfgang-Gaede-Str.\ 1, 76131 Karlsruhe, Germany}
\author{  Mark Sutton}
\affiliation{ Department of Physics and Astronomy, The University of Sussex, Brighton, BN1 9RH, United Kingdom }

\date{May 7, 2025}
%\date{December 30, 2024}

\begin{abstract}
   We determine the value of the strong coupling \as and study its running over a wide range of scales as probed by the dijet production process at hadron colliders, based on an NNLO QCD analysis of LHC dijet data.
  From a large subset of these data a value of $\asmz = 0.1178\,\pm\,0.0022$ is obtained for the strong coupling at the scale of the Z-boson mass \mz,
  using the invariant mass of the dijet system to select the scale where \as is probed. 
  The combination of different data sets enhances the reach and precision of the analysis in the multi-TeV range and allows for the first determination of \as  up to scales of 7\TeV.    
  Complementing the LHC data with dijet cross sections measured at the HERA electron--proton collider, the kinematic range is extended to test the
  running of the strong coupling towards smaller scales.
  Our results exhibit excellent agreement with predictions based on the renormalization group equation of QCD, and 
  represent a comprehensive test of the asymptotic behavior of QCD, spanning more than three orders of magnitude in energy scale.
\end{abstract}

\maketitle

% ---------------------------------------------------------------------
% INTRODUCTION
% ---------------------------------------------------------------------
\section{\label{sec:intro}Introduction}

The theory of  Quantum Chromodynamics (QCD)~\cite{Fritzsch:1973pi,Gross:1973ju,Gross:1973id,Politzer:1973fx,Gross:2022hyw} so far
successfully describes the dynamics and asymptotic behavior of the strong 
interaction. The renormalization group equation (RGE) of QCD predicts the scale evolution (``running'') of its
 coupling \as.
Consequently, the determination of the
strong coupling at different energy scales probes the non-Abelian gauge structure of QCD. Despite its outstanding importance as the only free parameter
of massless QCD, the value of the strong coupling constant at the reference scale of the Z-boson mass, \asmz is known with an uncertainty of approximately 1\,\%~\cite{ParticleDataGroup:2024cfk} and hence is one of the least precisely determined fundamental constants in physics.

In this letter we make use of new precise predictions from perturbative QCD (pQCD) for dijet production at next-to-next-to leading order (NNLO) including subleading color contributions~\cite{Czakon:2019tmo,Chen:2022tpk,Chen:2022clm}
to determine the value of the strong coupling constant \asmz.
We use precise dijet production data recorded by the
ATLAS~\cite{ATLAS:2013jmu,ATLAS:2017ble} and CMS~\cite{CMS:2012ftr,CMS:2017jfq,CMS:2023fix} experiments in proton--proton collisions ($pp$) at the LHC at center-of-mass energies of 7, 8, and 13\TeV.  The analysis is further extended to include dijet cross sections measured in electron--proton ($ep$) collisions at the HERA collider~\cite{H1:2000bqr,H1:2010mgp,ZEUS:2010vyw,H1:2014cbm,H1:2016goa}, which operated at considerably lower center-of-mass
energies of $\sqrt{s}=300$ and 320\GeV. This allows the investigation of the running of the strong coupling \asmur over energy scales ranging from a few GeV to the TeV regime.
The first theoretical studies of dijet production were performed at next-to-leading order (NLO) in pQCD in Refs~\cite{Kunszt:1992tn,Giele:1995kb}. 
The first applications of NNLO predictions to determine \as were carried out using $e^+e^-$ event shape data in Ref.~\cite{Dissertori:2007xa}  and using DIS jet production in Ref.~\cite{H1:2017bml}.
Recent determinations of \as in $pp$ collisions were performed at NNLO in a leading-color approximation with inclusive jet and dijet cross
sections~\cite{Britzger:2022lbf,CMS:2021yzl,CMS:2023fix,CMS:2024tux}, in aN$^3$LO~\cite{Cridge:2024exf}, and with multijet transverse energy correlations based on 3-jet NNLO
predictions~\cite{ATLAS:2023tgo,Czakon:2021mjy,Alvarez:2023fhi}. Extending to 
4.2~TeV, these predictions allow the measurement of \as at the 
largest scales attained up till now. 
By using multiple dijet data sets, our analysis achieves 
a considerably higher reach and resolution above
scales of one TeV,  allowing a measurement 
of \as with unprecedented precision in the range beyond 1\TeV, extending as far as 7~TeV.

% ---------------------------------------------------------------------
%                           METHODOLOGY
% ---------------------------------------------------------------------
\section{\label{sec:methodology}Methodology}
The value of \asmur is determined by performing a least-squares minimization of the complete NNLO pQCD predictions for selected inclusive dijet cross-sections from the ATLAS and CMS experiments at $pp$ center-of-mass energies $\sqrt{s}$ 
of 7, 8, and 13\TeV, summarized in Table~\ref{tab:datasets}.
\begin{table}[thb!]
%\footnotesize
%\begin{ruledtabular} % fill the entire width
 \resizebox{1.0\columnwidth}{!}{
  \begin{tabular}{lcccc} % lcdr
    \toprule\toprule
    \textrm{Data} &
    \textrm{$\sqrt{s}$ [TeV]} &
    \textrm{d$\sigma$} &
    $R$ & % _{\text{anti}-\kt}
    \textrm{$\mathcal{L}$}\\
    \midrule
    ATLAS  \cite{ATLAS:2013jmu} & 7  & $\frac{\text{d}^2\sigma}{\text{d}\mjj\text{d}\ys}$  & 0.6 &  4.5\,fb$^{-1}\pm1.8\,\%$\\
    CMS    \cite{CMS:2012ftr}   & 7  & $\frac{\text{d}^2\sigma}{\text{d}\mjj\text{d}\ym}$  & 0.7 &  5.0\,fb$^{-1}\pm2.2\,\%$\\
    CMS    \cite{CMS:2017jfq}   & 8  & $\frac{\text{d}^3\sigma}{\text{d}\ptavg\text{d}\ys\text{d}\yb}$  & 0.7 & 19.7\,fb$^{-1}\pm2.6\,\%$ \\
    ATLAS \cite{ATLAS:2017ble} & 13 & $\frac{\text{d}^2\sigma}{\text{d}\mjj\text{d}\ys}$  & 0.4 &  3.2\,fb$^{-1}\pm2.1\,\%$\\
    CMS    \cite{CMS:2023fix}   & 13  & $\frac{\text{d}^2\sigma}{\text{d}\mjj\text{d}\ym}$  & 0.8 & 33.5\,fb$^{-1}\pm 1.2\,\%$ \\
    CMS    \cite{CMS:2023fix}   & 13  & $\frac{\text{d}^3\sigma}{\text{d}\mjj\text{d}\ys\text{d}\yb}$  & 0.8 & 29.6\,fb$^{-1}\pm 1.2\,\%$ \\
    \bottomrule\bottomrule
  \end{tabular}
  } % end resizebox
  \caption{Selected dijet data sets with center-of-mass energy $\sqrt{s}$, cross-section definition $d\sigma$, jet size parameter $R$ and integrated luminosity $\mathcal{L}$.}
  \label{tab:datasets}
%\end{ruledtabular}
\end{table}
Two measurements from ATLAS  at $\sqrt{s}=$ 7
%~\cite{ATLAS:2013jmu}
and 13\TeV
%~\cite{ATLAS:2017ble}
are available as
functions of the dijet mass $\mjj = \sqrt{(p_{j_1}+p_{j_2})^2}$, and half of the absolute rapidity separation $\ys = \left|y_1-y_2\right|/2$, where $p_{j_1}, p_{j_2}$ and
$y_1, y_2$ denote the four-momenta and rapidities, respectively, of the two jets leading in \pt.
Double-differential measurements have been performed by CMS at $\sqrt{s}=$  
$7$
%~\cite{CMS:2012ftr}
and $13$\TeV,
%~\cite{CMS:2023fix}
as functions of \mjj and the maximum absolute rapidity, \ym,
of either of the two leading \pt jets.
CMS has also published triple-differential cross sections at $\sqrt{s}=$ 8
%~\cite{CMS:2017jfq}
and $13$\TeV
%~\cite{CMS:2023fix}
 as functions of either \mjj or the average transverse momentum of the two leading jets, \ptavg, half of their rapidity separation \ys, and the longitudinal boost of the dijet system given by $\yb = \left|y_1+y_2\right|/2$.
These measurements employ the anti-\kt jet algorithm~\cite{Cacciari:2008gp}, but use different jet size parameters $R$.
When cross sections are provided for more than one value of $R$,  the larger jet size parameter is selected due to the expected improved perturbative convergence~\cite{Currie:2018xkj}.
The CMS 13\TeV data are provided in both double- and triple-differential forms, but only one of the two data sets can be considered in the combined
study because of their experimental correlations. We choose the double-differential variant in the following due to its larger range in \mjj.
In order to reduce the sensitivity to parton distribution functions (PDFs), the selected data are restricted to $\ys<2.0$ (respectively $\ym<2.0$) and $\yb<1.0$, thereby excluding asymmetric parton configurations, where one parton carries a much smaller momentum fraction $x$ than the other. As an example, in the 8\,TeV data, this selection effectively restricts the PDFs to $x>10^{-2}$~\cite{Gehrmann-DeRidder:2019ibf}.
The selected data then have further experimental advantages since in the selected regions the tracking detectors of the experiments can be used.
Finally, altogether 367 out of 493 cross section measurements are considered in the \as determination.

% ---------------------- predictions -------------------
The dijet data are confronted with predictions in the framework of pQCD at NNLO~\cite{Currie:2017eqf,Gehrmann-DeRidder:2019ibf} as implemented in the NNLOJET framework~\cite{Ridder:2016rzm,Gehrmann:2018szu,Huss:2025iov}.
The \as sensitivity in this calculation arises from two components: the hard matrix elements and the PDFs.
The NNLO predictions include the full set of contributions, in particular all sub-leading color parts~\cite{Czakon:2019tmo,Chen:2022tpk,Chen:2022clm}, which are, for the first time, used in the determination of \as with LHC jet data.
Using the APPLfast library~\cite{,Britzger:2019kkb,Britzger:2022lbf}, the NNLO pQCD coefficients are stored independently of the \asmz value and PDF. The statistical uncertainty, derived from the Monte Carlo integration in NNLOJET, is typically around a percent or below. %, and is considered in the least-squares minimization.
The momentum distribution of partons inside the incoming proton is obtained from PDFs.
The $x$-dependence of the PDFs is defined at a starting scale $\mu_0$, and the PDFs are  evolved to the factorization scale $\muf$ using DGLAP evolution,
with \as as a free parameter, where they are convolved with the hard coefficients.
We set the scale $\mu_0$ to 90\GeV, a characteristic hard scale, and the $x$-dependence is taken from the PDF4LHC21 PDF combination~\cite{PDF4LHCWorkingGroup:2022cjn}.
The predictions further include bin-wise correction factors for non-perturbative
effects (NP) and higher-order electroweak (EW) contributions~\cite{Dittmaier:2012kx}.
Both correction factors and their uncertainties are taken as published by the experimental collaborations~\cite{ATLAS:2013jmu,CMS:2012ftr,CMS:2017jfq,ATLAS:2017ble,CMS:2023fix}.
Further details on the evaluation of the theory predictions are collected in Appendix~\ref{app:sec:theory}.
A comprehensive study to assess the agreement between the NNLO pQCD predictions and the dijet data, as well as the 
consistency of individual data sets across different kinematic regions and between multiple data sets is provided in Ref.~\cite{SupplementaryMaterial}.
Overall, good agreement is observed between the predictions and the data in all kinematic regions and for all data sets, with a very good consistency between the data sets. 

% ---------------------- fit algorithm -------------------
The value of \asmz is then determined through a least-squares fit of the NNLO predictions to the dijet data, similar to the method used in Refs.~\cite{H1:2017bml,Britzger:2019kkb}.
% ---------------------- uncertainties -------------------
The uncertainties considered in the fit include experimental, non-perturbative (NP), NNLO statistical,
and PDF uncertainties. Their covariance matrices also take correlations between data points and data sets into account.
Henceforth, the linearly propagated uncertainty from that fit will be denoted as ``(fit,PDF)'' uncertainty to emphasize that this uncertainty comprises experimental and PDF related uncertainties together.
Details on the \chisq minimization and considerations on the PDF uncertainties are discussed in Appendix~\ref{app:sec:fitalgo}. 

% PDFset, PDFas, PDFmu0
The predictions are associated with further uncertainties, related to the value of \asmz as used in the PDF determination and to the starting scale of the PDF evolution $\mu_0$. 
Since a variation in the factorization scale and \asmz are  related in the DGLAP framework, we apply the following approach. 
The starting scale is chosen corresponding to the bulk of the collider data entering into the PDF determination. 
A variation of \asmz in the original PDF determination is then expected to be mirrored by a $\mu_0$ variation, and vice-versa.
To validate this, we study additional fits, where the choice of the starting scale $\mu_0$ is varied by factors of 0.5 and 2, or where we select PDFs that were determined with values of \asmz varied by $\pm0.001$. We find that these variations yield very similar uncertainties.
In order to represent these two theoretical uncertainty components, we report half of the difference between two fits with $\mu_0$ varied by factors of 0.5 or 2 as an uncertainty (denoted  as ``($\mu_0$)'').

% scale uncertainty
An additional scale uncertainty accounts for missing higher orders beyond NNLO and for the actual choice of the renormalization \mur and factorization \muf scales.
Since the uncertainty associated with this choice is theoretical in nature and reflects the sensitivity of the prediction to unphysical scale choices, it cannot be constrained by data and is therefore not included in the \chisq function of the fit.
It is derived by varying $\mur$ and $\muf$ independently by factors of 0.5, 1, or 2 around the central value $\mur=\muf=\mjj$ in the complete NNLO pQCD predictions, 
omitting the two variations of $(0.5,2)$ and $(2,0.5)$, i.e.\ using the so-called 7-point scale variations. Half of the difference between the largest and smallest prediction is reported as \emph{scale} uncertainty (denoted as ``(\mur,\muf)''), since
the asymmetry in these variations is typically small. 

% ---------------------------------------------------------------------
% RESULTS
% ---------------------------------------------------------------------
\section{Results from LHC dijets}
\label{sec:resultslhc}

\begin{figure}[thbp!]
  \includegraphics[width=0.9\columnwidth,trim={10 20 0 20},clip]{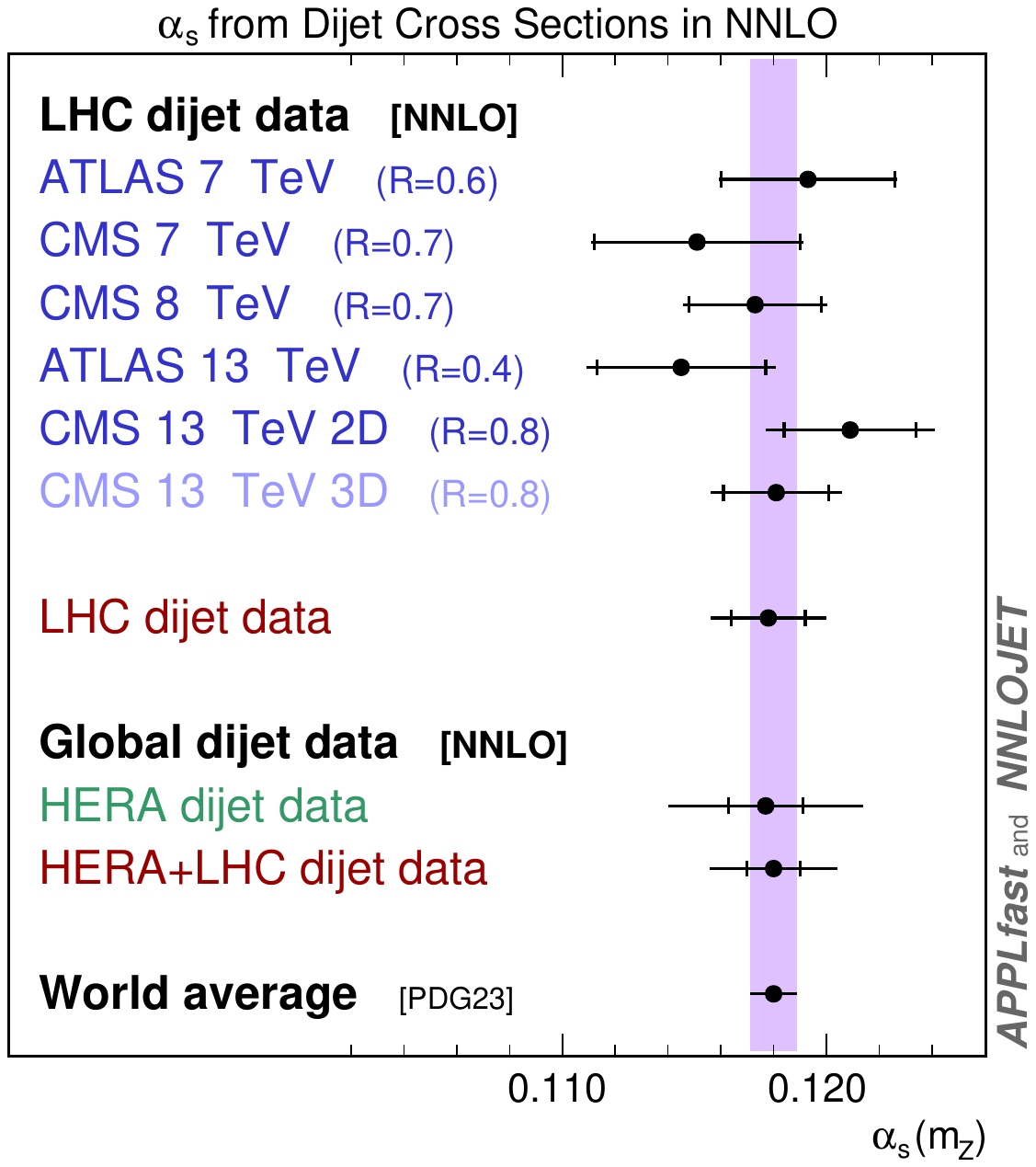} % \linewidth
  \caption{Comparison of \asmz determinations from dijet cross sections to the world average value.  The inner error bars indicate the (fit,PDF) uncertainty, and
    the outer error bars further include the scale and $\mu_0$ uncertainty.}
  \label{fig:alphas}
\end{figure}

The value of the strong coupling at the scale \mz~\cite{ParticleDataGroup:2024cfk} is determined from the five LHC dijet data sets
using complete NNLO pQCD predictions.
The fit exhibits an excellent consistency with $\chisq/\ndf=0.92$ and the value of \asmz
is determined to be
\begin{equation}
  \asmz = 0.1178~(14)_\text{(fit,PDF)}
  ~(1)_{(\mu_0)}
  ~(17)_{(\mur,\muf)}\,.
  \nonumber
\end{equation}
Fits of \asmz were also performed for individual data sets.  The results are collected in Table~\ref{tab:asmzresults} and displayed in Fig.~\ref{fig:alphas}, where they are compared to the combined fit from
all five data sets and to the world average value~\cite{ParticleDataGroup:2024cfk}.

\begin{table}[thbp!]
%  \footnotesize
  \begin{ruledtabular} % fill the entire width
%   \resizebox{1.0\columnwidth}{!}{
    \begin{tabular}{l@{~~}c@{~~~~}c} % lcdr
%      \toprule\toprule
      Data set  & \chisq/\ndf & \asmz \\
      \midrule
      ATLAS 7\TeV                   &    74.7/ 77 & $0.1193\,(33)\,( 4)\,(~\,6)$ \\ %  $(34)_\text{tot}$   $\chisq/\ndf=0.97$
      ATLAS 13\TeV                  &    87.7/106 & $0.1145\,(32)\,( 4)\,(16)$   \\ %  $(36)_\text{tot}$   $\chisq/\ndf=0.83$
      CMS 7\TeV                     &    50.7/ 45 & $0.1151\,(39)\,( 1)\,(~\,9)$ \\ %  $(41)_\text{tot}$   $\chisq/\ndf=1.13$
      CMS 8\TeV                     &    37.0/ 56 & $0.1173\,(25)\,( 1)\,(11)$   \\ %  $(28)_\text{tot}$   $\chisq/\ndf=0.66$
      CMS 13\TeV (2D)               &    71.6/ 78 & $0.1209\,(25)\,( 2)\,(20)$   \\ %  $(32)_\text{tot}$   $\chisq/\ndf=0.92$
      CMS 13\TeV (3D)               &   137.7/112 & $0.1181\,(20)\,( 1)\,(15)$   \\ %  $(25)_\text{tot}$   $\chisq/\ndf=1.23$
      \midrule                                                                   
      LHC dijets {\tiny (CMS13-2D)} &   335.3/366 & $0.1178\,(14)\,( 1)\,(17)$   \\ %  $(22)_\text{tot}$   $\chisq/\ndf=0.92$
      LHC dijets {\tiny (CMS13-3D)} &   397.9/400 & $0.1172\,(14)\,( 1)\,(14)$   \\ %  $(20)_\text{tot}$   $\chisq/\ndf=0.99$
      \midrule                                                                   
      HERA                          &    92.8/118 & $0.1177\,(14)\,( 1)\,(34)$   \\ %  $(36)_\text{tot}$   $\chisq/\ndf=0.79$
      LHC+HERA  {\tiny (CMS13-2D)}  &   428.4/485 & $0.1180\,(10)\,( 1)\,(22)$   \\ %  $(31)_\text{tot}$   $\chisq/\ndf=0.88$
      LHC+HERA  {\tiny (CMS13-3D)}  &   491.0/519 & $0.1177\,(10)\,( 1)\,(24)$   \\ %  $(29)_\text{tot}$   $\chisq/\ndf=0.95$
      %      \bottomrule\bottomrule
    \end{tabular}
%    }  % end resizebox
    \caption{\label{tab:asmzresults} Results of \asmz from fits of complete NNLO pQCD predictions to dijet cross section data. 
    Listed are the values of \asmz with the (fit,PDF) uncertainty, the $(\mu_0)$ uncertainty, and the scale uncertainty (\mur,\muf).
    The upper rows display results from fits to individual data sets. 
    The middle rows show results from fits to all studied LHC dijet data; once using the double-differential (2D) or triple-differential (3D) CMS 13\TeV data.
    The bottom rows show results from fits to HERA dijet data and from fits to LHC and HERA dijet data taken together.
    }
  \end{ruledtabular}
\end{table}

The results from the individual data sets  exhibit (fit,PDF) uncertainties in the range between $\pm0.0020$ to $\pm0.0039$.
Data sets with larger integrated luminosity or at higher center-of-mass energy yield smaller uncertainties.
The \asmz values are consistent with the world average value. 
It is observed that the determination of \asmz using all five LHC dijet data sets benefits significantly from independent measurements, extended
kinematic ranges, and multiple center-of-mass energies. Hence, the experimental uncertainties are found to
be reduced in the combined determination in comparison to any individual data set. An additional variant of the nominal fit  using the CMS 3D data is presented in appendix~\ref{appendix:CMS3D}.

% ---------------------------------------------------------------------
% RESULTS with HERA data
% ---------------------------------------------------------------------
\section{Including HERA dijet data}
The analysis is extended by further including data for dijet production in neutral-current deep-inelastic scattering (NC DIS)
taken at the HERA $ep$ collider.
These data, from the H1~\cite{H1:2000bqr,H1:2010mgp,H1:2016goa,H1:2014cbm} and ZEUS~\cite{ZEUS:2010vyw} collaborations,
have previously been used for \as determinations at NNLO accuracy~\cite{H1:2017bml,H1:2021xxi,ZEUS:2023zie} using the complete NNLO pQCD
predictions~\cite{Currie:2016ytq,Currie:2017tpe,Britzger:2019kkb}.
Further details on the data are collected in Appendix~\ref{sec:HERA} and the results of a fit to the HERA dijet measurements alone is presented in Table~\ref{tab:asmzresults}.
Using the HERA data provides competitive (fit,PDF) uncertainties in \asmz, but the fit exhibits a sizable scale dependence.

\begin{table}% [tbh!]
\begin{ruledtabular} % fill the entire width
  \begin{tabular}{lcc} % lcdr
    %  \begin{xtabular*}{\columnwidth}{lcc} % lcdr
    %  \begin{longtable}[e]{lcc} % lcdr
    %\toprule
    $\mur^\text{avg}$ {\scriptsize [GeV]}
    & \multicolumn{1}{c}{\asmz}
    & \multicolumn{1}{c}{\asmur}
    \\
    \midrule
      \phantom{11}7.4  & $0.1214\,(28)\,(1)\,(66)$   & 0.2013\,(82)\,(4)\,(196)      \\ %  mu=7.400000     
      \phantom{1}10.1  & $0.1207\,(15)\,(1)\,(53)$   & 0.1840\,(37)\,(2)\,(130)       \\ %  mu=10.119300     
      \phantom{1}13.3  & $0.1171\,(15)\,(0)\,(37)$   & 0.1654\,(31)\,(0)\,(77)         \\ %  mu=13.266500     
      \phantom{1}17.2  & $0.1151\,(20)\,(0)\,(26)$   & 0.1530\,(36)\,(1)\,(47)         \\ %  mu=17.233700     
      \phantom{1}20.1  & $0.1160\,(20)\,(1)\,(27)$   & 0.1498\,(34)\,(1)\,(46)         \\ %  mu=20.124600     
      \phantom{1}24.5  & $0.1159\,(18)\,(0)\,(23)$   & 0.1442\,(29)\,(1)\,(37)         \\ %  mu=24.494900     
      \phantom{1}29.3  & $0.1175\,(23)\,(0)\,(22)$   & 0.1418\,(33)\,(0)\,(32)         \\ %  mu=29.325800     
      \phantom{1}36.0  & $0.1171\,(26)\,(0)\,(24)$   & 0.1362\,(35)\,(1)\,(33)         \\ %  mu=35.916600     
      \phantom{1}49.0  & $0.1157\,(26)\,(1)\,(16)$   & 0.1275\,(31)\,(1)\,(20)         \\ %  mu=48.989800     
      \phantom{1}77.5  & $0.1105\,(37)\,(3)\,(12)$ & 0.1131\,(39)\,(3)\,(12)        \\ %  mu=77.459700     
      \phantom{1}250   & $0.1180\,(15)\,(1)\,(14)$   & 0.1025\,(11)\,(1)\,(11)         \\ %  mu=250.000000    
      \phantom{1}370   & $0.1181\,(15)\,(1)\,(16)$   & 0.0975\,(10)\,(1)\,(11)         \\ %  mu=370.000000    
      \phantom{1}550   & $0.1174\,(15)\,(1)\,(19)$   & 0.0925\,( 9)\,(1)\,(12)         \\ %  mu=550.000000    
      \phantom{1}810   & $0.1173\,(15)\,(2)\,(21)$   & 0.0885\,( 9)\,(1)\,(11)         \\ %  mu=810.000000    
                1175   & $0.1171\,(16)\,(2)\,(23)$   & 0.0848\,( 8)\,(1)\,(12)         \\ %  mu=1175.000000   
                1760   & $0.1171\,(17)\,(2)\,(25)$   & 0.0813\,( 8)\,(1)\,(12)         \\ %  mu=1760.000000   
                2545   & $0.1171\,(18)\,(2)\,(27)$   & 0.0783\,( 8)\,(1)\,(12)         \\ %  mu=2545.000000   
                3490   & $0.1171\,(20)\,(2)\,(29)$   & 0.0760\,( 8)\,(1)\,(12)         \\ %  mu=3490.000000   
                4880   & $0.1185\,(31)\,(3)\,(34)$   & 0.0742\,(12)\,(1)\,(13)         \\ %  mu=4880.000000   
                7040   & $0.1232\,(128)\,(12)\,(37)$ & 0.0734\,(43)\,(4)\,(13)         \\ %  mu=7042.000000   
                %\bottomrule
\end{tabular}
\end{ruledtabular}
  \caption{\label{tab:finalresults1column}
    Results for the running of the strong coupling.
    The values are reported
    for different \mur intervals.
    The columns show the central \mur value, the resulting value of \asmz, and the corresponding value of $\asmur$.
    The brackets denote the (fit,PDF), the $(\mu_0)$  and the  $(\mur,\muf)$ uncertainty.
  }
\end{table}

A combined fit to LHC and HERA dijet data is performed by considering all data and their uncertainties in the $\chisq$ function. Correlations between the dijet $ep$ and $pp$ processes arise from the PDF uncertainties.
Since the scale dependence is specific to each process, the scale uncertainty is evaluated in separate fits by applying the 7-point scale variation procedure independently for either the  $pp$ or the $ep$ calculation.
The full scale uncertainty is then obtained by adding quadratically the scale uncertainties derived individually for each process.

In the combined fit, altogether 612 dijet cross section data values are available.
After applying the \ys, \yb and \ym data selection criteria, 486 data points remain for the combined fit, which yields
\begin{equation}
  \asmz = 0.1180\,(10)_\text{(fit,PDF)}
  \,(1)_{(\mu_0)}
  \,(22)_{(\mur,\muf)}
  \nonumber
  \label{eq:asBoth}
\end{equation}
with $\chisq/\ndf=0.88$.  The \chisq value suggests an excellent consistency between the HERA and LHC data, as well as an outstanding agreement between
data and the NNLO pQCD predictions.  The \as value is found to be in excellent agreement with the world average value of \asmz of
0.1180\,(10)~\cite{ParticleDataGroup:2024cfk}.  As expected, the experimental uncertainties are reduced in the combined fit as compared to the fits to HERA or
LHC data alone.  
The individual scale uncertainty from the $pp$ NNLO calculations is found to be $\pm 0.0011$, and from $ep$ it amounts to $\pm 0.0019$. 
While each of these scale uncertainties is smaller than what was obtained for the separate fits to the individual data sets ($\pm 0.0017$ for $pp$ and $\pm 0.0034$ for $ep$, see Table~\ref{tab:asmzresults}), 
the resulting total
scale uncertainty is larger than in the fit to LHC data alone.

% -------------------------------------------------------- %
% Running
% -------------------------------------------------------- %
\section{\label{sec:running}Running of the strong coupling}
\begin{figure}[tb!]
  \includegraphics[width=.9\columnwidth]{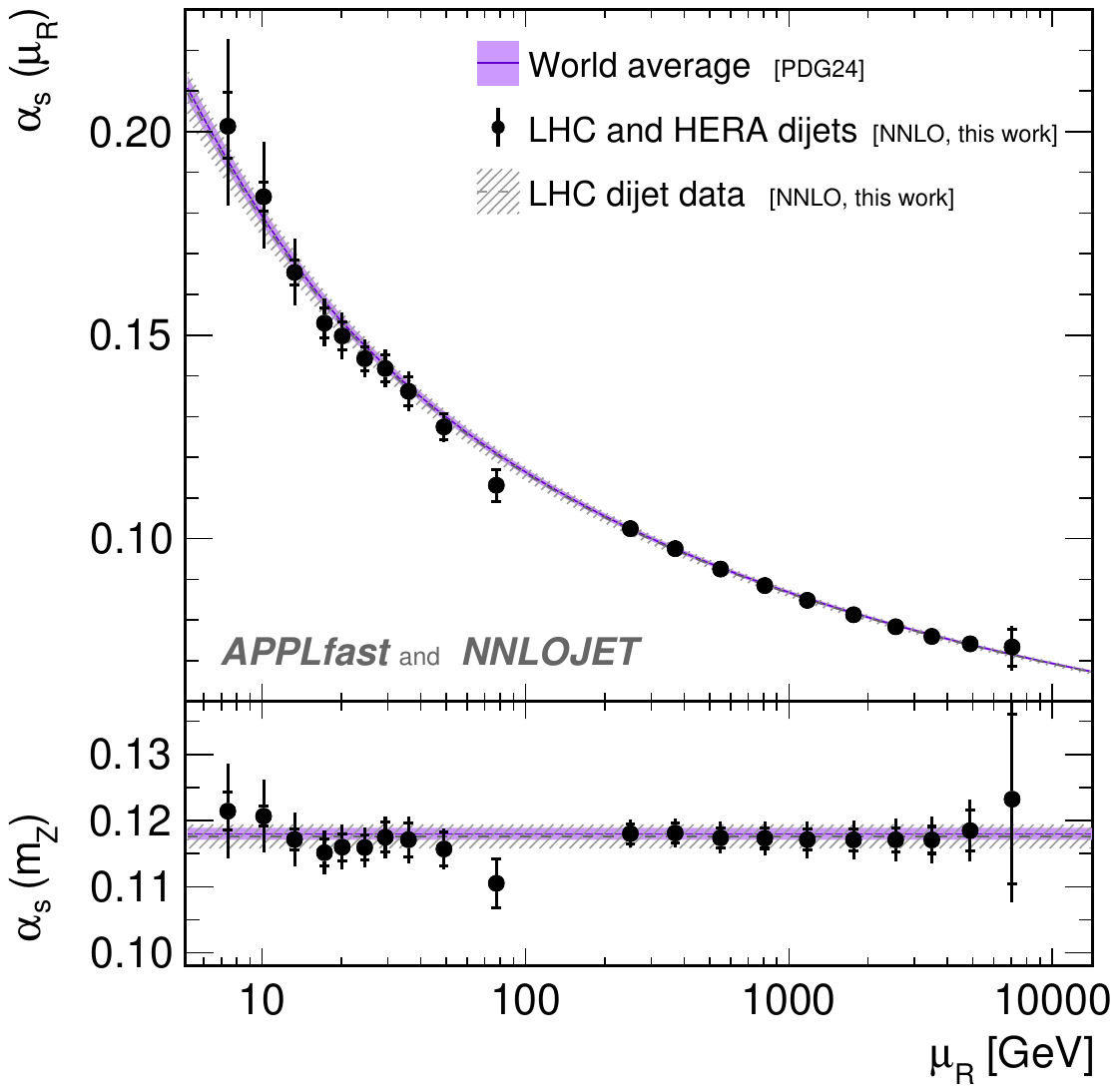} % \linewidth
  \caption{Running of the strong coupling as a function of the chosen renormalization scale.  The inner error bars indicate the
    (fit,PDF) uncertainty, and the outer error bars the total uncertainty.  The upper panel displays the values \asmur and the lower panel displays
    the respective \asmz value and the world average value~\cite{ParticleDataGroup:2024cfk}.  The hatched area indicates the value
    of \asmz from LHC dijet data and its running as a function of \mur.}
  \label{fig:runningV1}
\end{figure}
The asymptotic behavior of the strong coupling is one of the key properties of QCD~\cite{Gross:1973ju,Gross:1973id,Politzer:1973fx}.  Its prediction
needs to be validated with experimental data, for example by probing the running of \asmur by determining \as at different values
of \mur.  For such a study, dijet cross sections represent a particularly powerful opportunity, since the dijet system provides a natural choice
for the renormalization scale \mur, which in principle could be chosen freely.  As before, for dijet production in $pp$ collisions  \mur is identified with
\mjj, while for $ep$ data $\mur^2=\Qsq+\ptavg^2$~\cite{H1:2017bml} is used. The \mur values of the HERA and LHC dijet cross sections span over three orders of magnitude from about
7\GeV up to 7\TeV.

Each cross section measurement is then assigned a single representative value of \mur.  These values are used (only) to group the data into 20
distinct \mur intervals.  It is confirmed that in each \mur interval, data from multiple data sets are considered.

We then perform a single fit to all dijet data, where, for each of the individual ranges of \mjj, a separate \asmz value is used for the prediction.
In this fit, the assumption of the QCD running enters in each interval only within a very limited range, and in the evolution
of the PDFs from $\mu_0$ to $\muf$ (using $\mu_0=90\GeV$ and $\muf=\mur$).
The technical fit parameter of \asmz in each interval is evolved to the appropriate scale value \asmur as needed for the computation of the NNLO prediction.
The advantage of a single fit to determine multiple \asmz values at a time, in comparison to an alternative approach where each value is determined in a
separate fit~\cite{H1:2017bml,ATLAS:2023tgo}, is that the inference benefits from constraints on the correlated experimental uncertainties, as well as
on the PDF uncertainties.
In addition, the uncertainties in the resulting \asmz values have known correlations and these values can therefore be used
in further analyses.
It has to be noted, that the lowest \mur interval needs to be considered with some care, since these data are below the $2m_b$ threshold,
and thus our computations in the five flavor number scheme are at the edge of their validity. 
However, it is found that these data do not impact other data in the fit, which is also seen from the resulting weak correlations,
and thus this result can be neglected also at a later stage.
The result at $\mur=7.4\GeV$ is therefore reported here for completeness as in previous analyses~\cite{H1:2017bml,Britzger:2019kkb}.

\begin{figure}[tb!]
  \includegraphics[width=.9\columnwidth]{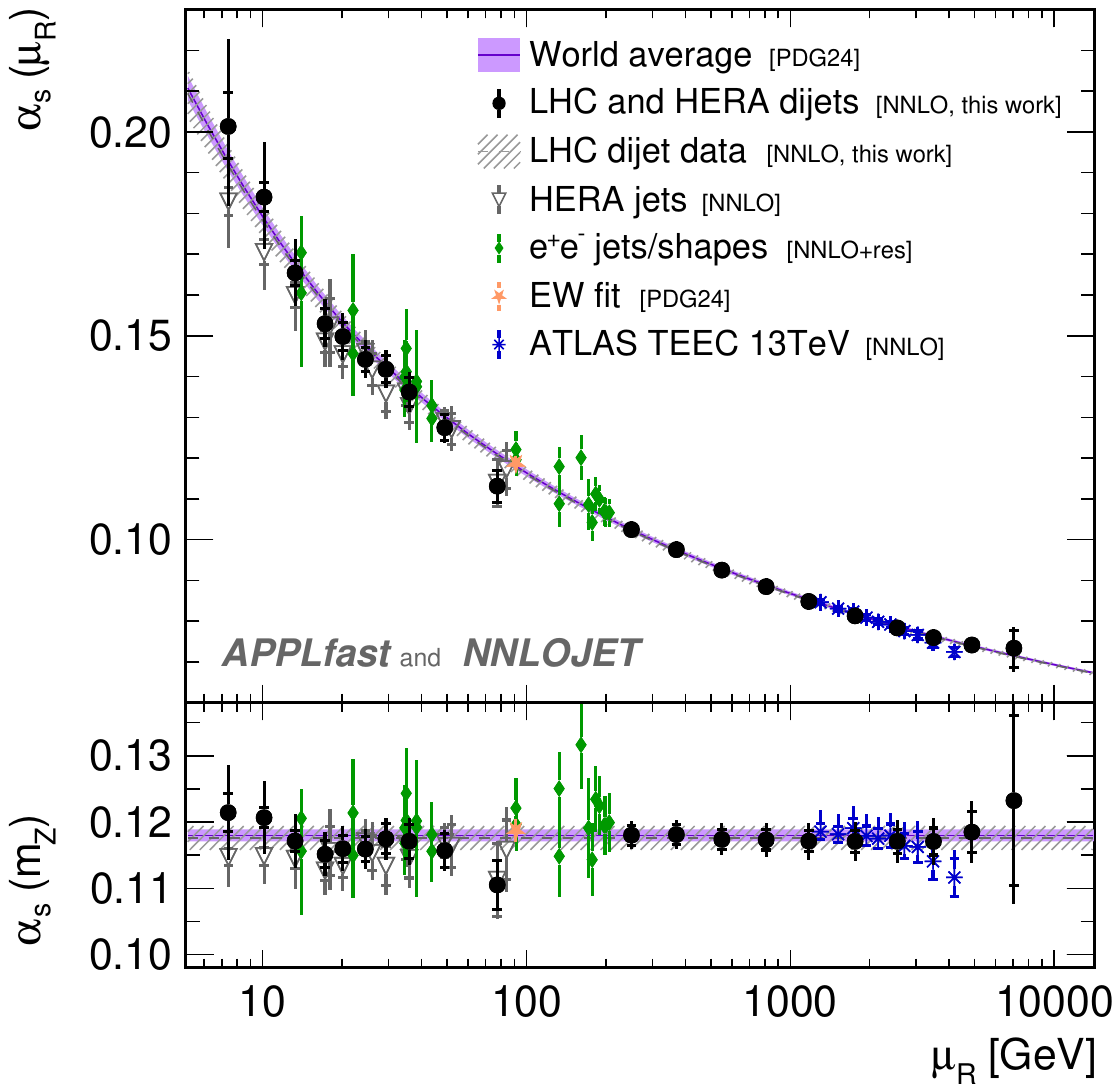} % \linewidth
  \caption{ Running of the strong coupling as a function of the chosen renormalization scale.  The inner error bars indicate the
    (fit,PDF) uncertainty, and the outer error bars the total uncertainty.  The upper panel displays the values \asmur and the lower panel displays
    the respective \asmz value and the world average value~\cite{ParticleDataGroup:2024cfk}.  The shaded area indicates the value
    of \asmz from LHC dijet data and its running as a function of \mur.
  }
  \label{fig:runningV2}
\end{figure}

The results from this single fit are presented in Table~\ref{tab:finalresults1column} and
the related correlations of the (fit,PDF) uncertainty are listed in Appendix~\ref{sec:tables}.
The results are compared to the expectation from the QCD RGE in Fig.~\ref{fig:runningV1}, where
in the lower panel the results of the 20 fit parameters for \asmz are displayed, while the upper panel shows the respective
values for \asmur.
The \asmz values are evolved to the central value of each \mur interval, illustrating the running of the strong coupling.
Overall, excellent agreement with the expectation from the RGE running (when using the world average value for \asmz) is observed over the
entire range from about 7\GeV up to 7\TeV.
At scales of about a few hundred \GeV, the size of the experimental and theoretical uncertainties are of similar size (about $\pm0.0015$), while in the \TeV regime the
experimental uncertainties dominate.
In Fig.~\ref{fig:runningV2} our results are further compared to \as extractions from inclusive jet and dijet data by the H1 and ZEUS collaborations at HERA~\cite{H1:2017bml,ZEUS:2023zie}, event shape observables at the PETRA or LEP $e^+e^-$ colliders~\cite{Dissertori:2009ik,Bethke:2009ehn,Schieck:2012mp,OPAL:2011aa}, a result from a global electroweak fit~\cite{ParticleDataGroup:2024cfk} and measurements of energy--energy correlations in $pp$ collisions by ATLAS at the LHC~\cite{ATLAS:2023tgo}.
Our results exhibit significantly smaller uncertainties and cover a significantly larger range in scale than any previous determination of \asmur.

% ---------------------------------------------------------------------
% SUMMARY
% ---------------------------------------------------------------------
\section{\label{sec:summary}Summary} %

We have determined the strong coupling $\asmz$ from dijet data for the first time based on complete NNLO pQCD predictions. Using LHC data collected by the ATLAS and CMS collaborations at center-of-mass energies of 7, 8, and 13 TeV the strong coupling is determined to be
\begin{equation}
  \asmz = 0.1178\,(22)_\text{(tot)}\,,
  \nonumber
\end{equation}
where experimental, PDF, and scale uncertainties are all of similar size.
This value is consistent with the world average.

% , while for the HERA case the scale uncertainties are dominant. 
Including dijet cross sections measured in
electron--proton collisions at the HERA collider,
makes this one of the most comprehensive and precise tests of the QCD renormalization group running of $\as(\mu)$ to date.
The running is probed by a fit to individual \mjj ranges, and excellent agreement is found with the running predicted by QCD.
Through the inclusion of both HERA and LHC data, the behavior of the strong coupling as a function of energy is tested over an unprecedented range, from about 7\GeV to 7\TeV.
The presented results significantly improve our knowledge of the strong coupling in the \TeV regime compared to previous determinations.

\section*{\label{sec:Note}Note added}
Recently, the CMS Collaboration has released a determination of \as and its running in the range $103\GeV<\mur<1600\GeV$ using inclusive jet data at the LHC at various $\sqrt{s}$~\cite{CMS:2024rkg} in addition to HERA DIS data. Their determination make use of NNLO pQCD predictions in the leading-color approximation. Their results are in agreement with ours.

\begin{acknowledgments}
  This research was supported in part by the German Federal Ministry of Education and Research (BMBF) under grant number 05H21VKCCA, by the UK Science
  and Technology Facilities Council, by the Swiss National Science Foundation (SNF) under contracts 200021-197130 and 200020-204200, by the National Science Foundation of China (grant No.12475085 and No.12321005), by the Research
  Executive Agency (REA) of the European Union through the ERC Advanced Grant MC@NNLO (340983) and ERC Advanced Grant TOPUP (101019620) and by the Funda\c{c}\~{a}o para a Ci\^{e}ncia e
  Tecnologia (FCT-Portugal), through the programatic funding of R\&D units reference UIDP/50007/2020 and under project CERN/FIS-PAR/0032/2021. CG and MS were supported by the IPPP Associateship program
  for this project.
  This work was performed in parts on the HoreKa supercomputer funded by the Ministry of Science, Research and the Arts Baden-Württemberg and by the
  BMBF. We also thank the bwHPC, CERN, DESY and MPCDF computing facilities for providing computational resources.
\end{acknowledgments}

% ---------------------------------------------------------------------
% REFERENCES
% ---------------------------------------------------------------------
%\clearpage
\bibliography{apssamp}% Produces the bibliography via BibTeX.

% ---------------------------------------------------------------------
% APPENDIX
% ---------------------------------------------------------------------
%\clearpage
\onecolumngrid
\vspace{4em}
\begin{center}
\large
\textbf{Appendix}
\end{center}
\twocolumngrid
\appendix

% ---------------------------------------------------------------------
%                           METHODOLOGY
% ---------------------------------------------------------------------
\section{\label{app:sec:theory}Theory predictions}
The pQCD cross section for the process with two initial-state hadrons is obtained from the factorization formula as the convolution of the
PDFs of the incoming protons and the hard scattering cross section
\begin{align}
  \text{d}\sigma =
  \sum_{a,b}
  \int
  \frac{\text{d}x_1}{x_1}
  \frac{\text{d}x_2}{x_2}
  f_a(x_1,\muf)
  f_b(x_2,\muf)
  \text{d}\hat{\sigma}_{ab}(\mur,\muf)\,,
  \nonumber
\end{align}
where $f_a(x,\muf)$ denotes the density of the partons of type $a$ in the incoming proton at the factorization scale $\muf$ carrying the longitudinal
momentum fraction $x$.
Both contributions are sensitive to the value of $\alpha_s$, as
\begin{align}
  \text{d}\hat{\sigma}_{ab}(\mu) &\equiv \text{d}\hat{\sigma}_{ab}(\mu,\alpha_s(\mu))
                                   ~~\text{and}~~
  \label{eq:pqcd}\\
  f_a(x,\mu) &\equiv f_a(x,\mu,\alpha_s(\mu)) \,.
\end{align}
The \as\ dependence in the partonic cross section is explicit through the perturbative expansion, which for dijet production up to NNLO reads
\begin{align}
  \text{d}\hat{\sigma}_{ab}(\as) =&
                                    \left(\tfrac{\as(\mu)}{2\pi}\right)^2\text{d}\hat{\sigma}_{ab,\text{LO}}
                                    +
                                    \left(\tfrac{\as(\mu)}{2\pi}\right)^3\text{d}\hat{\sigma}_{ab,\text{NLO}}
                                    \nonumber\\
                                  &+
                                    \left(\tfrac{\as(\mu)}{2\pi}\right)^4\text{d}\hat{\sigma}_{ab,\text{NNLO}}
                                    +
                                    \mathcal{O}(\as^5(\mu))\,.
                                    \label{eq:sighat}
\end{align}
The value of $\as(\mu)$ is obtained from \asmz from the renormalization group running in the modified minimal subtraction ($\overline{\text{MS}}$) scheme, i.e.\ $\as(\mu)=\alpha_{\text{s},\overline{\text{MS}}}^{(5)}(\mu,\asmz)$, in three-loop order~\cite{Tarasov:1980au,Larin:1993tp} as implemented in CRunDec~\cite{Schmidt:2012az}.
The evolution is performed with with $n_f=5$ active flavors throughout, in particular also beyond the top-quark mass threshold.
This is consistent with the perturbative calculation that does not include top-quark effects and thus effectively treats the top quark in the decoupling limit.
The evolution of the PDFs with respect to a scale $\mu$ is governed by the DGLAP equations, whose splitting kernels $\mathcal{P}$
depend on $\as(\mu)$,
\begin{equation}
  \mu^2\tfrac{\text{d}f}{\text{d}\mu^2} = \mathcal{P(\as)}\otimes f\,.
\end{equation}
The $x$-dependence of the PDFs can be fixed at a starting scale $\mu_0$ with value $f_{\mu_0}$, and subsequently evolved to a scale $\mu$ using the DGLAP evolution
\begin{equation}
  f_a(x,\mu,\as) = ({\Gamma}(\mathcal{P},\mu,\mu_0,\as) \otimes f_{\mu_0})_a\,,
\end{equation}
where $\Gamma$ denotes the DGLAP kernels which are evaluated at three-loop order~\cite{Vogt:2004mw,Moch:2004pa} using the program
Apfel++~\cite{Bertone:2013vaa,Bertone:2017gds}. % and validated with QCDNUM~\cite{Botje:2010ay}.
We set the scale $\mu_0$ of the evolution to
$90\GeV$ and the $x$-dependence of $f_{\mu_0,x}$ is taken from PDF4LHC21~\cite{PDF4LHCWorkingGroup:2022cjn}. %is taken from the PDF4LHC21 combination of global PDF determinations~\cite{PDF4LHCWorkingGroup:2022cjn}.
The NNLO cross section is obtained by integrating the dijet parton level predictions (Eq.\eqref{eq:sighat}) over the bin-dependent kinematic region $\Omega_i$, $\sigma_{\text{NNLO},i}=\int_{\Omega_i}\text{d}\sigma$, using the dijet parton level matrix elements and phase-space integration routines implemented in NNLOJET.
Our fit algorithm requires recalculating the predictions for different values of \asmz and corresponding PDFs. To streamline this, NNLOJET is interfaced with the APPLfast library
~\cite{Britzger:2019kkb,Britzger:2022lbf} which integrates the grid tools APPLgrid~\cite{Carli:2005,Carli:2010rw} and
fastNLO~\cite{Kluge:2006xs,Britzger:2012bs}.
The resulting interpolation grids for the dijet data sets typically have sub-permille accuracy.
The NNLO prediction is supplemented with additional correction factors to account for non-perturbative effects (NP) and higher-order electroweak (EW) contributions~\cite{Dittmaier:2012kx}, $c_{\text{NP}}$ and $c_{\text{EW}}$:
\begin{align}
  % use index 'i' for an individual bin?
  \sigma_i =
  c_{\text{NP},i}
  \cdot
  c_{\text{EW},i}
  \cdot
  \sigma_{\text{NNLO},i}\,.
  \label{eq:sigma}
\end{align}
Both correction factors are taken as published by the experimental collaborations~\cite{ATLAS:2013jmu,CMS:2012ftr,CMS:2017jfq,ATLAS:2017ble,CMS:2023fix}.
A consistent treatment of NP effects across all data sets is desirable but beyond the scope of this article.
Hence, different hadronization and parton-shower models are applied, reflecting variations in the Monte Carlo event generators~\cite{Sjostrand:2006za,Sjostrand:2007gs,Bahr:2008pv,Bellm:2015jjp} used to derive $c_{\text{NP},i}$.
Such variations are considered by the collaborations in the assignment of uncertainties.

% ---------------------------------------------------------------------
\section{\label{app:sec:fitalgo}Fit algorithm and uncertainties}
The objective function used in the fitting algorithm to determine the value of \asmz\ is
derived from normally distributed relative uncertainties and defined as~\cite{H1:2014cbm}
\begin{equation}
  \chisq = \sum_{i,j}
  \log\frac{\varsigma_i}{\sigma_i}
  \left(V_\text{exp} + V_\text{NP} + V_\text{NNLOstat} + V_\text{PDF}\right)_{ij}^{-1}
  \log\frac{\varsigma_j}{\sigma_j}\,,
  %\label{eq:chisq}
  \nonumber
\end{equation}
where the double-sum runs over all data points, $\varsigma_{i}$ denotes the measured cross section, $\sigma_i$ denotes the theory prediction.
The \chisq\ is minimized using TMinuit's Migrad algorithm~\cite{James:1975dr,Brun:1997pa}.
The covariance matrices $V_\text{exp}$, $V_\text{NP}$, $V_\text{NNLOstat}$, and $V_\text{PDF}$ represent the relative experimental, NP, NNLO statistical, and PDF uncertainties, respectively. 
% exp
The experimental uncertainties are reported by the experimental collaborations and account for many systematic sources as well as statistical components including correlations from unfolding.
Correlations between the experimental uncertainties of individual data sets are not provided and hence are assumed to be uncorrelated, which is certainly correct for the statistical components.
A recent report from CMS~\cite{CMS:2024tux} using inclusive jet data at different $\sqrt{s}$ 
indicates that the dominating uncertainty from jet energy calibration and resolution may be considered as uncorrelated
between such data sets, supporting that the omission of correlations is justified.
The non-perturbative correction uncertainties (cf. Sec. A) are provided by the experimental collaborations and are derived using different MC event generators and hadronization models, and from variations of their different respective model parameters. To represent both correlated and uncorrelated effects due to these various components, we follow the approach of Ref.~\cite{H1:2021xxi}, and assume a bin-to-bin correlation of 0.5.
The NNLO statistical uncertainties originate from the Monte Carlo integration in NNLOJET and are typically at the percent level or below.
The PDF uncertainties are obtained from the respective PDF set in the LHAPDF format~\cite{Buckley:2014ana}, and evaluated at $\mu_0$.
By considering them as a covariance matrix in \chisq, the PDF uncertainties are further constrained by the jet data. 
The PDFs carry further uncertainties due to differing theoretical assumptions, data selections, and inference methods imposed by the PDF fitting groups. 
In the PDF4LHC21 PDF set, however, such differences are already included in the uncertainty representation~\cite{PDF4LHCWorkingGroup:2022cjn} and represent differences between the MSHT~\cite{Bailey:2020ooq}, NNPDF3.1~\cite{Ball:2017nwa} and CT18~\cite{Hou:2019efy} PDFs. 
Dedicated fits using these different PDF sets confirm that the PDF uncertainty indeed covers such differences.
Results when using yet different PDFs, such as ABMP~\cite{Alekhin:2017kpj}, NNPDF4.0~\cite{NNPDF:2021njg}, or HERAPDF2.0~\cite{H1:2015ubc}, are typically found to be well within $2\sigma$ of the PDF
uncertainty.

% ---------------------------------------------------------------------
% Results with CMS 13 3D data
% ---------------------------------------------------------------------
\section{Fits using CMS 13\TeV triple-differential data}
\label{appendix:CMS3D}

The CMS Collaboration reported dijet cross sections at $\sqrt{s}=13\TeV$ 
also in triple-differential variants as a function of \ys, \yb, and \mjj or \ptavg~\cite{CMS:2023fix}. 
Besides observables and different binnings, the analyzed data and
experimental methods are equivalent in these three variants, and therefore these data sets cannot be used in a fit together because of their experimental correlations.
This section discusses the triple-differential measurement $\frac{\text{d}^3\sigma}{\text{d}\mjj\text{d}\ys\text{d}\yb}$
for a determination of \asmz instead of their double-differential variant (cf.\ Tab.~\ref{tab:datasets})
When restricting the data to $\ys<2.0$ and $\yb<1.0$, similar to the fits in Sec.~\ref{sec:methodology}, the fit to these data results in a value of $\chisq/\ndf$ of $1.23$ for 113 data points and provides
$\asmz=0.1181\,(20)_{(\text{fit,PDF})}\,( 1)_{(\mu_0)}\,(15)_{(\mur,\muf)}$. 
Using the triple-differential data as an alternative to the double-differential variant in the combined fit,
the value
\begin{equation}
  \asmz = 0.1172\,(14)_{(\text{fit,PDF})}\,(1)_{(\mu_0)}\,(14)_{(\mur,\muf)}
  \nonumber
\end{equation}
is derived with $\chisq/\ndf$ of $0.99$.
The result is in good agreement with that obtained when using the double-differential data.
For the main analysis presented in this letter, the double-differential CMS data is chosen rather than the triple-differential cross sections, as the sensitivity to the PDF parameters is lower, and the double-differential data reaches higher values of \mjj, while the sensitivity of the data to \asmz is similar.

% ---------------------------------------------------------------------
% RESULTS with HERA data
% ---------------------------------------------------------------------
\section{Including HERA dijet data}
\label{sec:HERA}

We extend our analysis by further including data for inclusive dijet production in neutral-current deep-inelastic scattering (NC DIS) reported by
the H1~\cite{H1:2000bqr,H1:2010mgp,H1:2016goa,H1:2014cbm} and ZEUS~\cite{ZEUS:2010vyw} collaborations, together with complete NNLO pQCD
predictions~\cite{Currie:2016ytq,Currie:2017tpe,Britzger:2019kkb}.
These data have already been used for \as determinations at NNLO accuracy~\cite{H1:2017bml,H1:2021xxi,ZEUS:2023zie}, and thus, the method and data selection from H1~\cite{H1:2017bml} is closely followed: four data sets at $\sqrt{s}=300$ and 320\GeV at lower or higher photon virtualities \Qsq being considered, and the fit methodology differing only in the choices for the PDF and $\mu_0$.
In addition, data from the ZEUS collaboration recorded at $\sqrt{s}=320\GeV$ and for $\Qsq>125\GeV^2$ are also included, similar to Refs.~\cite{H1:2021xxi,ZEUS:2023zie}.
All five data sets, summarized in Tab.~\ref{tab:HERAdata}, employ the \kt jet algorithm with $R=1.0$ and represent double-differential cross sections as a function of \Qsq and \ptavg.
\begin{table}[thbp!]
  %\begin{ruledtabular} % fill the entire width
   \resizebox{1.0\columnwidth}{!}{
    \begin{tabular}{lcccc} % lcdr
      \toprule\toprule
      Data set & $\sqrt{s}$ [GeV] & Cuts % & d$\sigma$ & $R$
      \\
      \midrule
      H1 300 GeV high-\Qsq\,\cite{H1:2000bqr}       & 300  & --                \\ % 
      H1 HERA-I low-\Qsq\,\cite{H1:2010mgp}         & 320  & $\mu>2m_b$        \\ % 
      H1 HERA-II low-\Qsq\,\cite{H1:2016goa}        & 320  & $\mu>2m_b$        \\ % 
      H1 HERA-II high-\Qsq\,\cite{H1:2014cbm}       & 320  & --                \\ % 
      ZEUS HERA-I+II high-\Qsq\,\cite{ZEUS:2010vyw} & 320  & $\ptavg>15\GeV$   \\ % 
      \bottomrule\bottomrule
    \end{tabular}
    } % end resizebox
    \caption{Summary of the HERA data sets for dijet production with the \kt jet algorithm with jet size parameter $R=1.0$.}
    \label{tab:HERAdata}
%\end{ruledtabular}
\end{table}
The ZEUS data are restricted to $\ptavg > 15\GeV$ to exclude infrared sensitive data points~\cite{Currie:2017tpe}.
At lower \Qsq, data points with a typical scale smaller than twice the bottom quark mass ($\mu<2m_b$) are excluded in the nominal fit, since the
predictions are performed with $n_f=5$~\cite{H1:2017bml}.
The correlations between data sets are described in Refs.~\cite{H1:2017bml,H1:2021xxi}.
The scales are identified with $\mur^2=\muf^2=\Qsq+\ptavg^2$.
From fits to individual data sets, consistent results are obtained for \chisq/\ndf and \asmz for the H1 data as in
Ref.~\cite{H1:2017bml}. For the ZEUS data a value of $\chisq/\ndf=11.8/15$ is obtained with 
$\asmz=0.1164\,(33)_\text{(fit,PDF)}\,(20)_{(\mur,\muf)}$.  
A fit to all HERA dijet data result in a value $\asmz = 0.1177\,(14)_\text{(fit,PDF)}\,(1)_{(\mu_0)}\,(34)_{(\mur,\muf)}$ with $\chisq/\ndf=92.8/118$.
As expected, these results are very similar to those reported from H1 data alone~\cite{H1:2017bml}, as the ZEUS
dijet data add only modestly to the sensitivity.
These results represent the first determination of \asmz at NNLO using only DIS dijet production, including data from H1 and ZEUS.
The value of \asmz as determined in a single fit to HERA and LHC dijet data taken together was reported in Table~\ref{tab:asmzresults} (cf.\ Sec.~\ref{sec:HERA}).
This analysis benefits from theory predictions for dijet production at NNLO and from
independent, and thus fully uncorrelated, experimental setups.
When the triple-differential data from CMS at 13\TeV are used instead of the double-differential variants in that fit, a value of \asmz of
\begin{equation}
0.1177\,(10)_{(\text{fit,PDF})}\,(1)_{(\mu_0)}\,(27)_{(\mur,\muf)}
\nonumber
\end{equation}
is obtained with $\chisq/\ndf$ of $0.95$ for 520 individual data points.
This result is in good agreement with that obtained using the double-differential data instead.

% ---------------------------------------------------------------------
% Tables?
% ---------------------------------------------------------------------
\section{\label{sec:tables}Resulting correlations}
%%%%%%%%%%%%%%%%%%%%%%%%%%%%%%%%%%%%%%%%%%%%%%%%%%%%%%%%%%%%%%%%%%%%%%%%%%%%%%%%%%%%%%%%%%%%%%%
%                        Compact table with correlations
%%%%%%%%%%%%%%%%%%%%%%%%%%%%%%%%%%%%%%%%%%%%%%%%%%%%%%%%%%%%%%%%%%%%%%%%%%%%%%%%%%%%%%%%%%%%%%%
\begin{table}
\scriptsize
%\begin{ruledtabular} % fill the entire width
\resizebox{1.0\columnwidth}{!}{
  \begin{tabular}{l@{~~}d@{\!}d@{\!}d@{\!}d@{\!}d@{\!}d@{\!}d@{\!}d@{\!}d@{\!}d@{\!}d@{\!}d@{\!}d@{\!}d@{\!}d@{\!}d@{\!}d@{\!}d@{\!}d@{\!}d} % lcdr
    \toprule\toprule
    $\mur$ 
    & \multicolumn{20}{c}{\multirow{2}*{Correlations}}
    \\
    {\tiny [GeV]} & & & \\    
    \midrule
      \phantom{11}7.4  & - & 56      & 29  & 21  & 19  & 22  & 15  & 17  & 16  & 12  &  2  &  1  & $-1$  & $-2$  & $-3$  & $-3$  & $-3$ & $-3$ & $-1$ & 0 \\
      \phantom{1}10.1  &           56 & -      & 65  & 50  & 49  & 50  & 37  & 38  & 36  & 23  &  9  &  8  &  5  &  2  &  0  & $-2$  & $-3$ & $-3$ & $-2$ & 0 \\
      \phantom{1}13.3  &           29 & 65      & -  & 58  & 52  & 54  & 40  & 45  & 39  & 23  & 11  & 11  &  9  &  7  &  5  &  2  &  1  & 0  & 0  & 1 \\
      \phantom{1}17.2  &           21 & 50      & 58  & -  & 48  & 52  & 39  & 44  & 41  & 24  &  9  &  9  &  8  &  7  &  5  &  3  &  2  & 1  & 1  & 1 \\
      \phantom{1}20.1  &           19 & 49      & 52  & 48  & -  & 52  & 39  & 38  & 41  & 24  &  9  &  9  &  9  &  8  &  7  &  5  &  4  & 2  & 1  & 1 \\
      \phantom{1}24.5  &           22 & 50      & 54  & 52  & 52  & -  & 55  & 49  & 53  & 36  & 10  & 11  & 11  & 10  &  9  &  7  &  5  & 3  & 2  & 1 \\
      \phantom{1}29.3  &           15 & 37      & 40  & 39  & 39  & 55  & -  & 41  & 44  & 33  &  6  &  8  &  9  & 10  &  9  &  8  &  7  & 5  & 3  & 1 \\
      \phantom{1}36.0  &           17 & 38      & 45  & 44  & 38  & 49  & 41  & -  & 39  & 28  &  5  &  6  &  8  &  8  &  8  &  8  &  7  & 5  & 3  & 1 \\
      \phantom{1}49.0  &           16 & 36      & 39  & 41  & 41  & 53  & 44  & 39  & -  & 31  &  4  &  5  &  6  &  7  &  8  &  7  &  6  & 5  & 3  & 1 \\
      \phantom{1}77.5  &           12 & 23      & 23  & 24  & 24  & 36  & 33  & 28  & 31  & -  &     &  1  &  2  &  2  &  3  &  4  &  4  & 3  & 2  & 1 \\
      \phantom{1}250   &            2 &  9      & 11  &  9  &  9  & 10  &  6  &  5  &  4  &  0  & -  & 90  & 87  & 83  & 78  & 71  & 64  & 54  & 36  &  9 \\
      \phantom{1}370   &            1 &  8      & 11  &  9  &  9  & 11  &  8  &  6  &  5  &  1  & 90  & -  & 95  & 91  & 87  & 80  & 72  & 61  & 40  & 10 \\
      \phantom{1}550   &         $-1$ &  5      &  9  &  8  &  9  & 11  &  9  &  8  &  6  &  2  & 87  & 95  & -  & 97  & 93  & 88  & 80  & 67  & 45  & 11 \\
      \phantom{1}810   &         $-2$ &  2      &  7  &  7  &  8  & 10  & 10  &  8  &  7  &  2  & 83  & 91  & 97  & -  & 97  & 93  & 86  & 74  & 49  & 12 \\
                1175   &         $-3$ &  0      &  5  &  5  &  7  &  9  &  9  &  8  &  8  &  3  & 78  & 87  & 93  & 97  & -  & 97  & 92  & 80  & 55  & 14 \\
                1760   &         $-3$ & $-2$     &  2  &  3  &  5  &  7  &  8  &  8  &  7  &  4  & 71  & 80  & 88  & 93  & 97  & -  & 96  & 87  & 62  & 17 \\
                2545   &         $-3$ & $-3$     &  1  &  2  &  4  &  5  &  7  &  7  &  6  &  4  & 64  & 72  & 80  & 86  & 92  & 96  & -  & 92  & 70  & 21 \\
                3490   &         $-3$ & $-3$     &  0  &  1  &  2  &  3  &  5  &  5  &  5  &  3  & 54  & 61  & 67  & 74  & 80  & 87  & 92  & -  & 78  & 27 \\
                4880   &         $-1$ & $-2$     &  0  &  1  &  1  &  2  &  3  &  3  &  3  &  2  & 36  & 40  & 45  & 49  & 55  & 62  & 70  & 78  & -  & 30 \\
                7040   &           0  &  0      &  1  &  1  &  1  &  1  &  1  &  1  &  1  &  1  &  9  & 10  & 11  & 12  & 14  & 17  & 21  & 27  & 30  & - \\
  \bottomrule\bottomrule       
  \end{tabular}
  } % end resizebox
  \caption{\label{tab:correlations}Correlations of the (fit,PDF) uncertainty from the fit of 20 \asmz parameters to HERA and LHC dijet data.
  }
%\end{ruledtabular}
\end{table}
%%%%%%%%%%%%%%%%%%%%%%%%%%%%%%%%%%%%%%%%%%%%%%%%%%%%%%%%%%%%%%%%%%%%%%%%%%%%%%%%%%%%%%%%%%%%%%%
The resulting correlations of the (fit,PDF) uncertainty in the combined fit of 20 parameters to the HERA and LHC dijet data
are listed in Table~\ref{tab:correlations}.
These correlations originate from the combined determination of 20 fit parameters and from correlated uncertainties between individual cross section values.
In the region where HERA or LHC data are important, \mur smaller or larger 100\GeV respectively, the correlations originate predominantly from correlated experimental systematic uncertainties.
Hence, correlations are found to be positive.
Correlations between low and high scales, respectively between HERA and LHC data, originate from PDF uncertainties.

The additional $(\mu_0)$ and $(\mur,\muf)$ uncertainties are fully correlated.

%\onecolumngrid
%\clearpage
%\twocolumngrid

% ---------------------------------------------------------------------
% Supplementary material
% ---------------------------------------------------------------------
\onecolumngrid
\clearpage
%\begin{widetext}
\begin{center}
\large
\textbf{Supplementary material}
\end{center}
%\section*{\label{sec:supplement}Supplementary material}

% - - - - - - - - - - - - - - - - - - - - - - - - - - - - - - - - - - -
%                  Consistency Study
% - - - - - - - - - - - - - - - - - - - - - - - - - - - - - - - - - - -
\subsection{Consistency study}
\label{sec:ConsistencyStudy}
Before determining the value of \asmz from dijet cross section measurements, we perform a study to investigate the agreement between the NNLO pQCD predictions and the data, and to test the
self-consistency of the individual data sets, as well as the consistency of multiple data sets together.

\subsubsection{Individual data sets}
\begin{figure*}[tbh!]
  \includegraphics[width=0.24\textwidth,trim={20 0 20 0},clip]{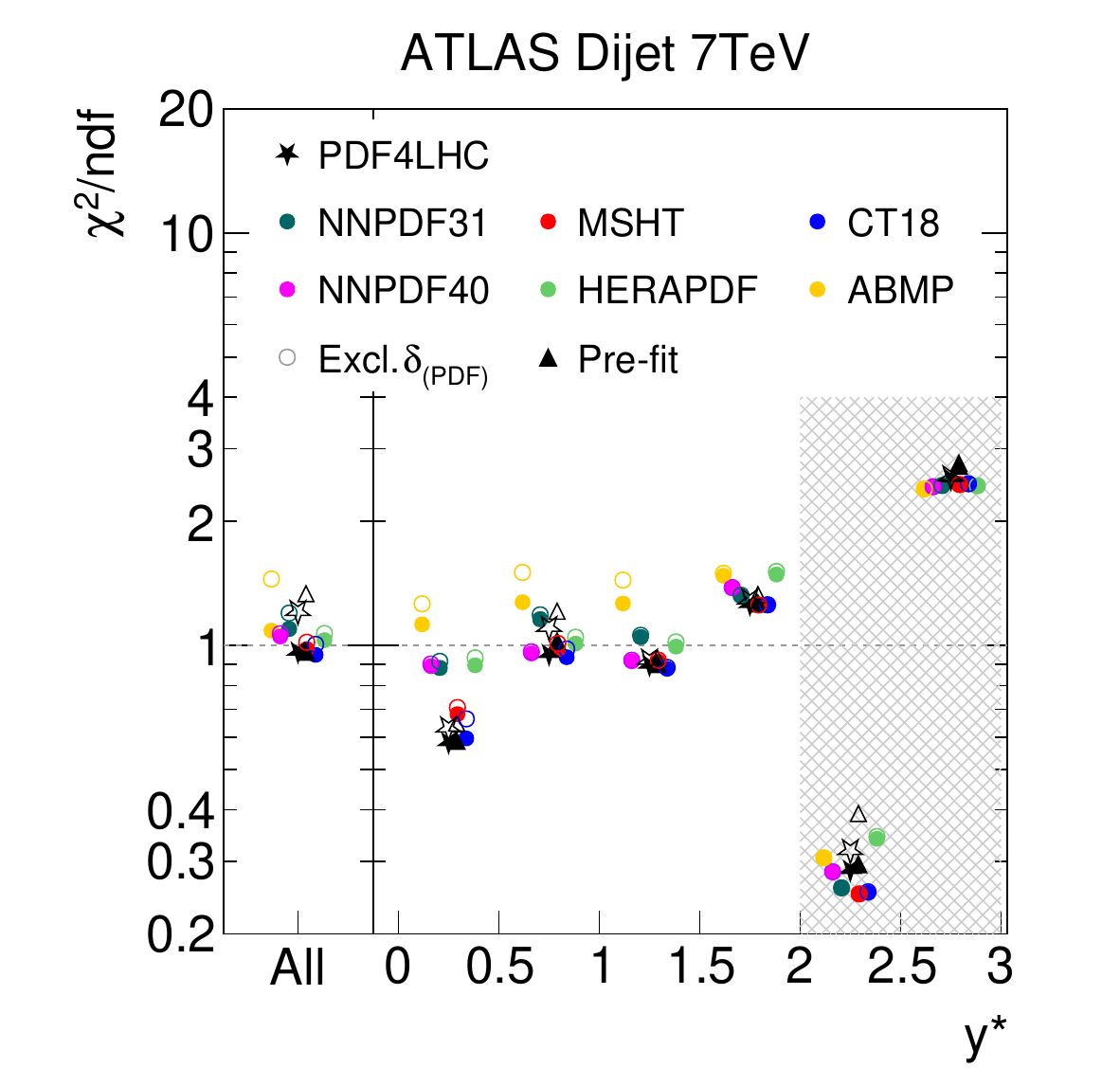}
  \includegraphics[width=0.24\textwidth,trim={20 0 20 0},clip]{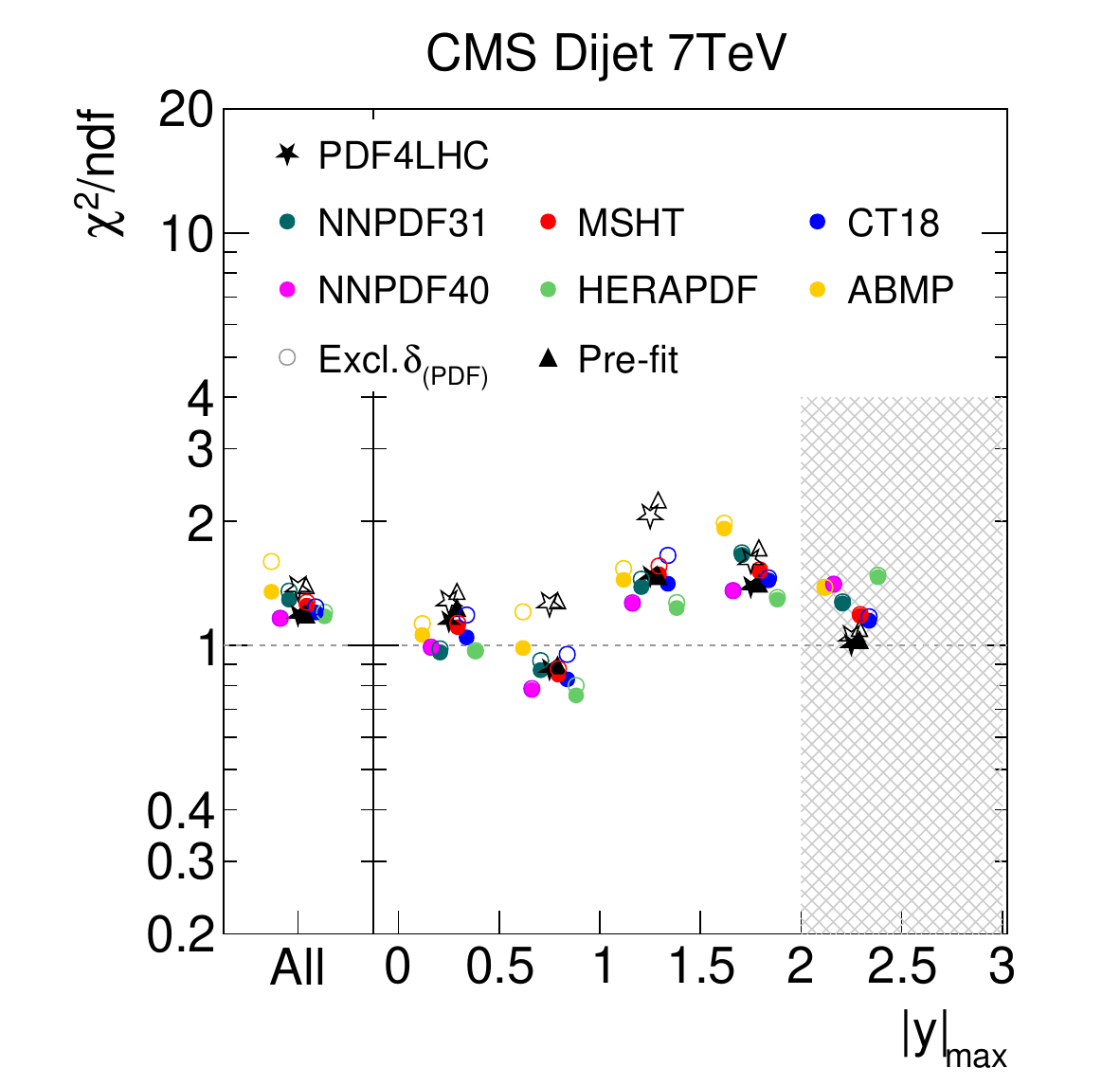}
  \includegraphics[width=0.24\textwidth,trim={20 0 20 0},clip]{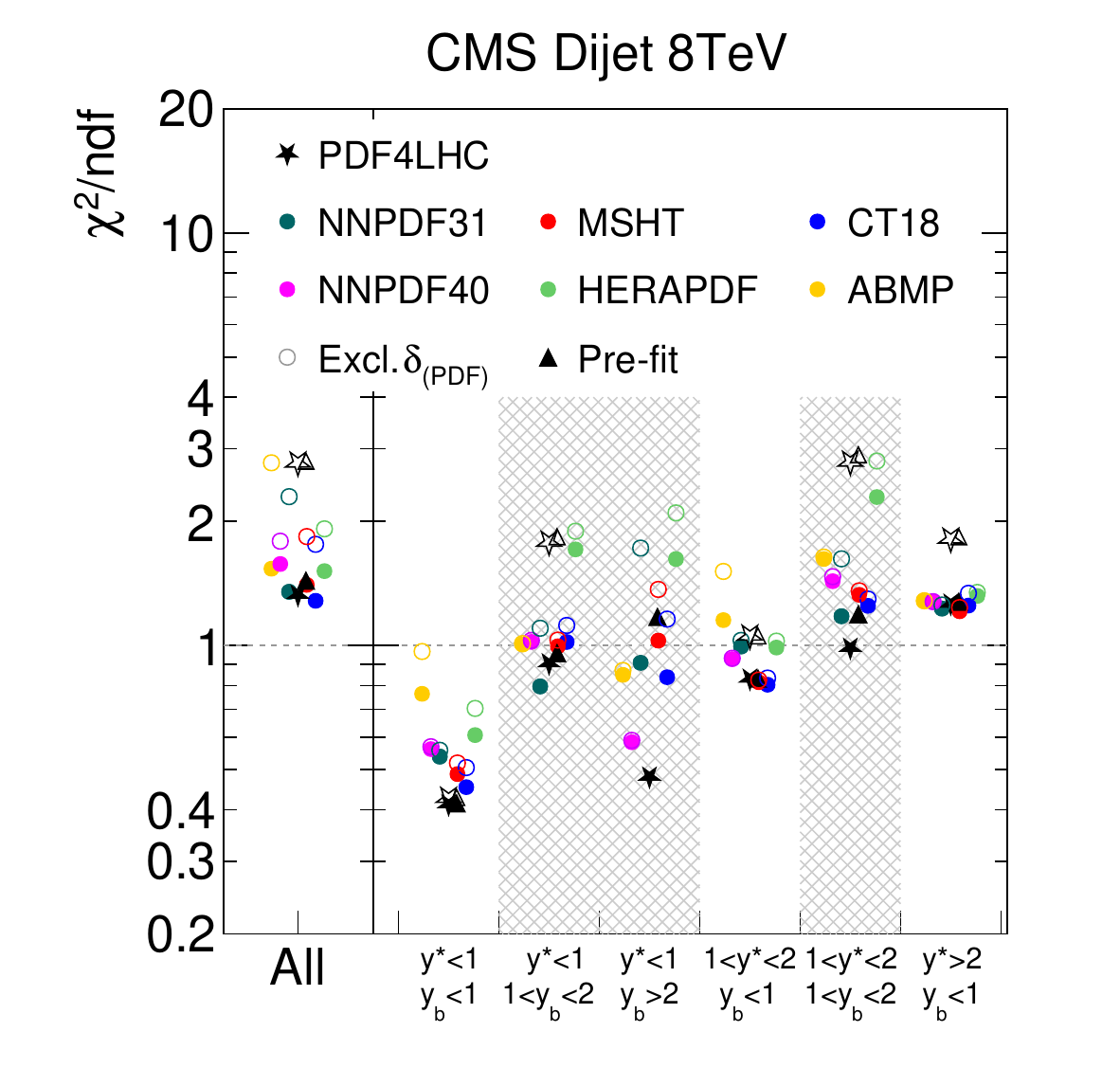}
  \includegraphics[width=0.24\textwidth,trim={20 0 20 0},clip]{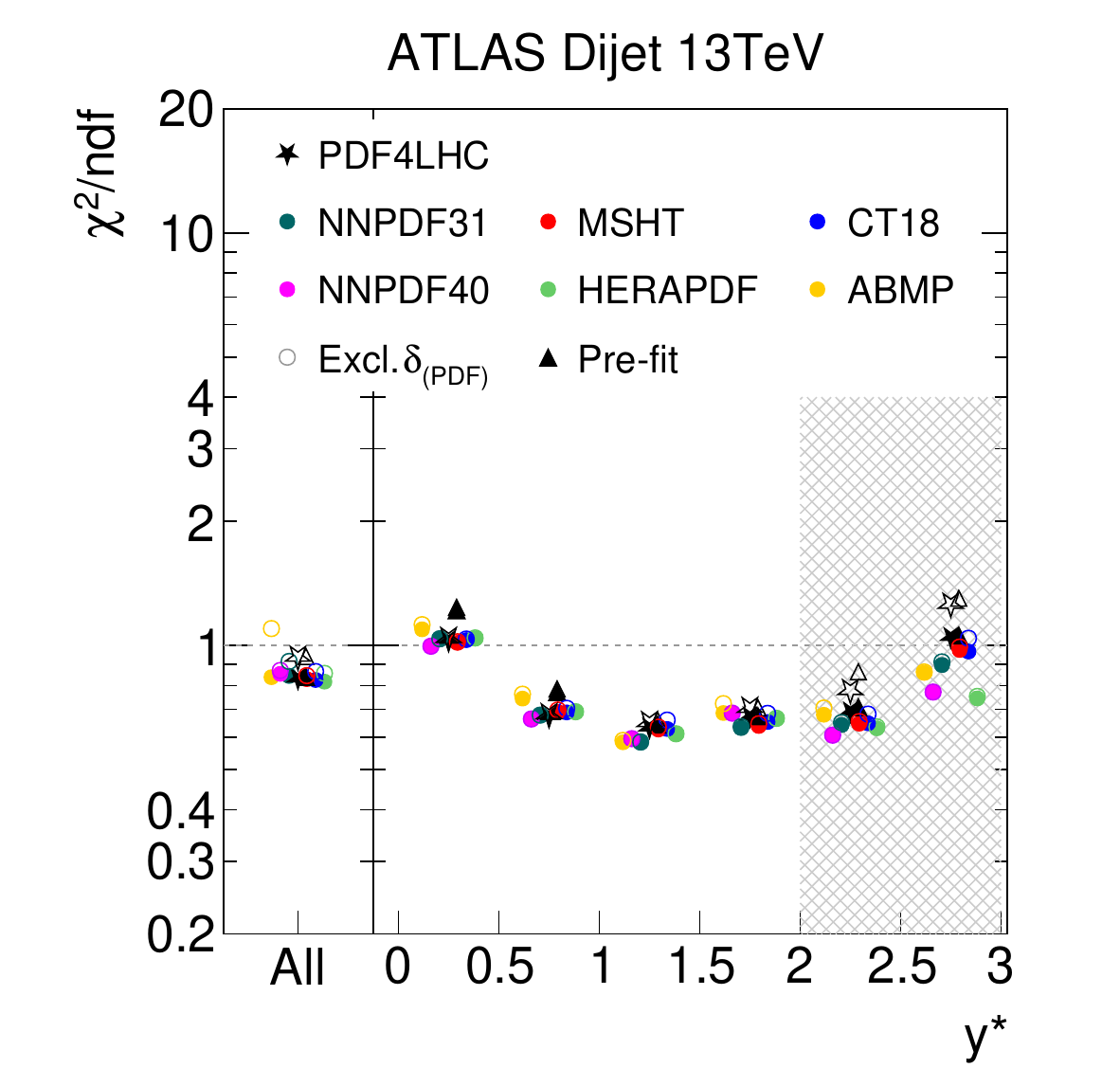}
  \includegraphics[width=0.24\textwidth,trim={20 0 20 0},clip]{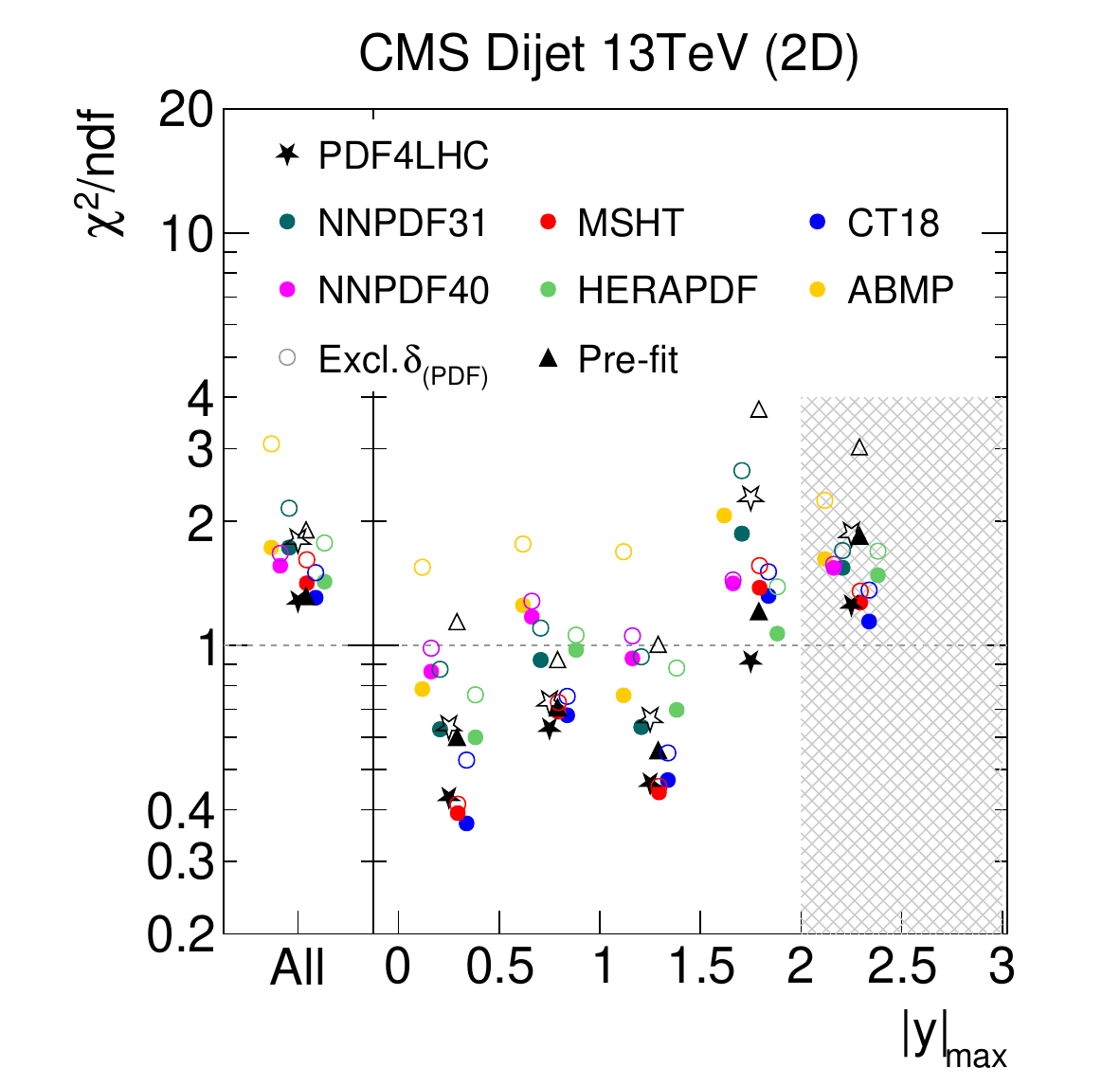}
  \includegraphics[width=0.24\textwidth,trim={20 0 20 0},clip]{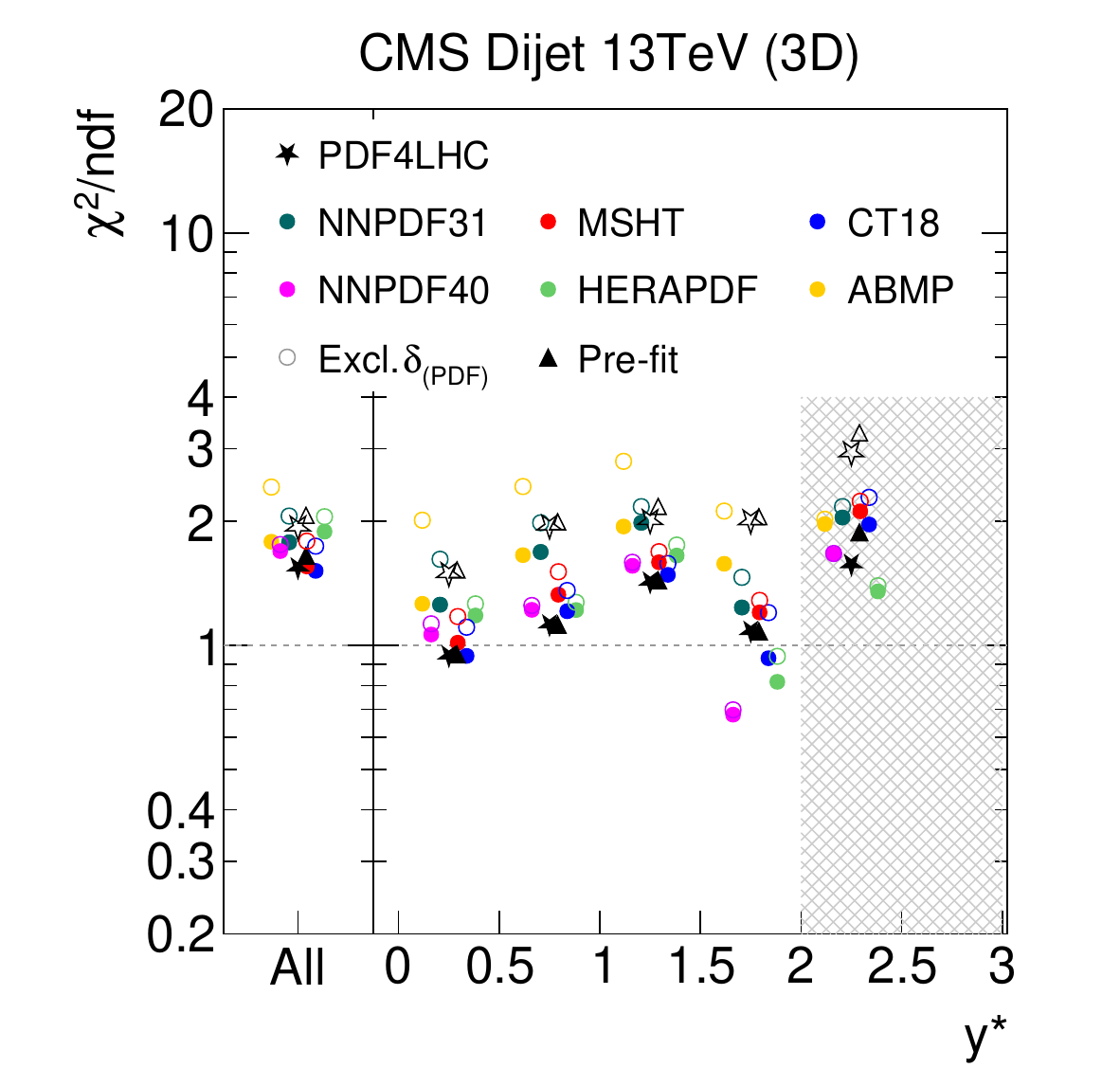}
  \includegraphics[width=0.24\textwidth,trim={20 0 20 0},clip]{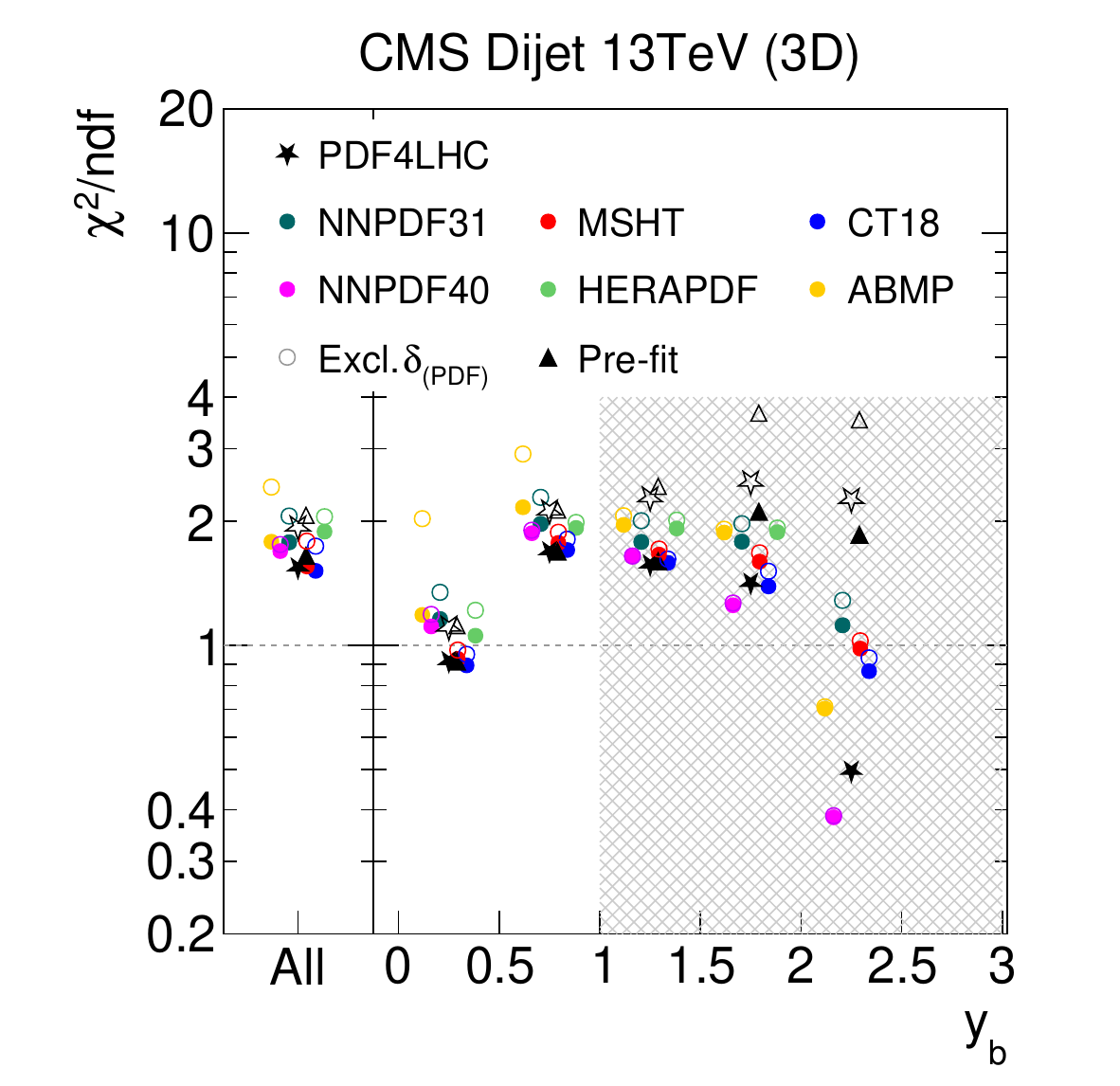}
  \includegraphics[width=0.24\textwidth,trim={20 0 20 0},clip]{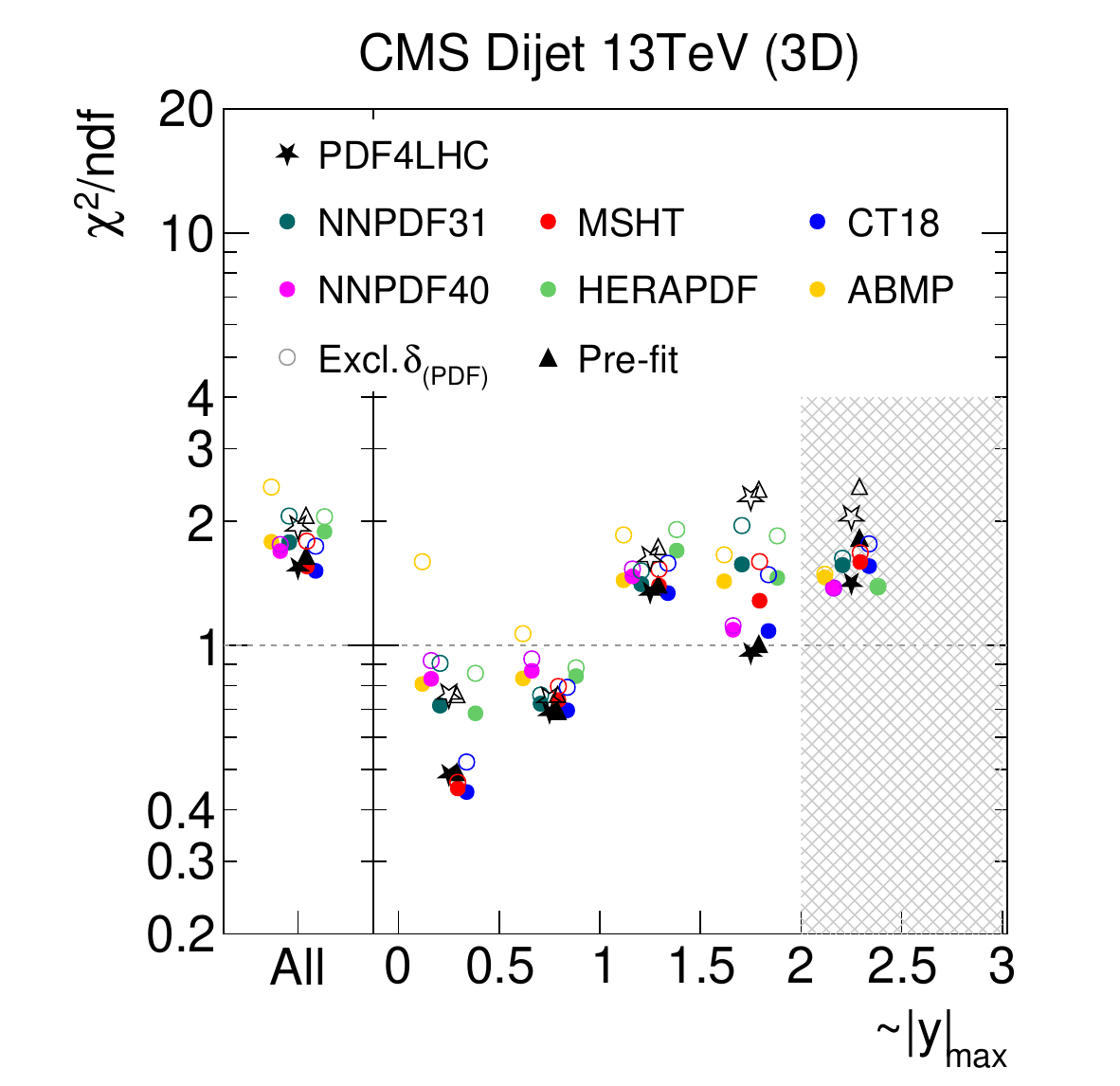}
  \caption{
    Post-fit \chisq/\ndf values of \asmz-fits to individual \ys or \yb-ranges of each data set, and values from fits to entire single
    data sets (denoted as \emph{All}).
    The top row shows dijet cross sections from ATLAS and CMS at 7\,TeV, CMS at 8\,TeV, and ATLAS at 13\,TeV.
    The bottom row shows the \chisq/\ndf values for the double-differential data from CMS at 13\,TeV (left), and three
    studies of the triple-differential CMS 13\,TeV data for individual \ys, \yb, or \ym ranges.
    The color coding indicates different PDF sets, as specified in the Panel. The colored markers are vertically displaced for better visibility.
    The open markers indicate post-fit values, where $V_\text{PDF}$ is not included in the \chisq-calculus, for each of the PDF set studied.
    The black triangle indicates the pre-fit value of the nominal NNLO pQCD predictions, when using the PDF4LHC21 PDF set.
    The shaded area indicates ranges, which are not included in the nominal combined fits (for the CMS 13\,TeV triple-differential
    data, there is some ambiguity due to the second $y$ cut, \ys\ or \yb, respectively).
    \label{fig:chisq}
    }
\end{figure*}

Determinations of \as are performed for individual \ys, \yb, \ym bins of the individual data sets.
Each double-differential data sets has five or six \ys (\yb, \ym) ranges, and these are studied separately in the following.
For the triple-differential CMS 8\TeV cross sections we study the six (\ys,\yb)-bins separately.
For the triple-differential CMS 13\TeV data, three studies are performed for individual \ys, \yb, or \ym ranges.
We consider the PDF sets PDF4LHC21~\cite{PDF4LHCWorkingGroup:2022cjn}, CT18~\cite{Hou:2019efy}, MSHT~\cite{Bailey:2020ooq},
NNPDF3.1~\cite{Ball:2017nwa}, NNPDF4.0~\cite{NNPDF:2021njg}, ABMP~\cite{Alekhin:2017kpj}, and HERAPDF2.0~\cite{H1:2015ubc}.
In addition, we study fits, where the PDF uncertainties are not considerd in the \chisq calculus (denoted as Excl. $\delta_\text{(PDF)}$).
The resulting values of \chisq/\ndf of these fits with a variety of PDF sets are displayed in FIG.~\ref{fig:chisq}.

It is observed that the pre-fits yield reasonable $\chisq$ values, indicating an initial good agreement between the NNLO predictions and the data.
Significant exceptions are only observed for very large values of \ys, \yb, or \ym, which
may be related to the increased PDF dependence in these kinematic regions, and either poorly determined PDFs or too tight PDF uncertainities.
It is further observed that the post-fit values of \chisq/\ndf yield reasonable values ranging from 0.29 to 2.5, while most of the values are around unity, i.e.\ in the range between 0.6 to 1.3.
%
%% -- different PDFs... we conclude...
The \chisq/\ndf values for the different PDF sets are reasonably consistent.
The values for ABMP and HERAPDF2.0 are slightly higher, which is expected, since these PDFs include few or no data from the LHC experiments.
The PDF4LHC21 PDF set shows good agreement with the data in all fits, and this PDF set has often one of the smallest \chisq/\ndf values of all PDF variants,
which supports the choice of PDF4LHC21 for our main result.
%
%% -- PDFs uncertainties included/excluded... we conclude...
In several fits, the \chisq/\ndf values without PDF uncertainty are somewhat larger than those with PDF uncertainties included, which indicates the importance of the PDF uncertainty in these bins.
%
%% -- chi2 of 'All' fits
The \chisq/\ndf values of the fits to all data of a single data set (\emph{All}) also yield reasonable values with \chisq/\ndf value ranging from 0.8 to 1.6.
However, for some data sets, these \chisq/\ndf values are somewhat larger than the ones obtained for individual $y$ ranges.
This may indicate some slight tension in these data and originate from the assumptions of the correlation model of the data systematic uncertainties, or from PDFs.

%% ---  nuisance parameters of syst. uncertainties.
In conclusion, we observe, that the NNLO predictions provide an overall good description of the data and are suitable for an unbiased determination
of \asmz.
For our nominal fits, we impose cuts on $\ys<2.0$ and $\yb<1.0$ to reduce the PDF sensitivity and reduce some moderate tensions within certain data sets.

\subsubsection{Multiple data sets analysed together}
To assess the consistency between the individual data sets, \as fits are performed considering data points from all data sets. Since the CMS 13\TeV data are provided in both double- and triple-differential forms, but only one of the two data sets can be included in this combined study due to their statistical correlations, we perform the study twice, once for each data set.

The various data sets are provided for distinct \ys or \yam ranges, and we define three intervals in the following: $0\leq\ys<1$, $1\leq\ys<2$,
$\ys\geq2$ (the $\yam$-ranges from Ref.~\cite{CMS:2012ftr} are interpreted as \ys for this particular study).
\begin{figure}[tbh!]
  \includegraphics[width=0.24\columnwidth,trim={20 0 20 0},clip]{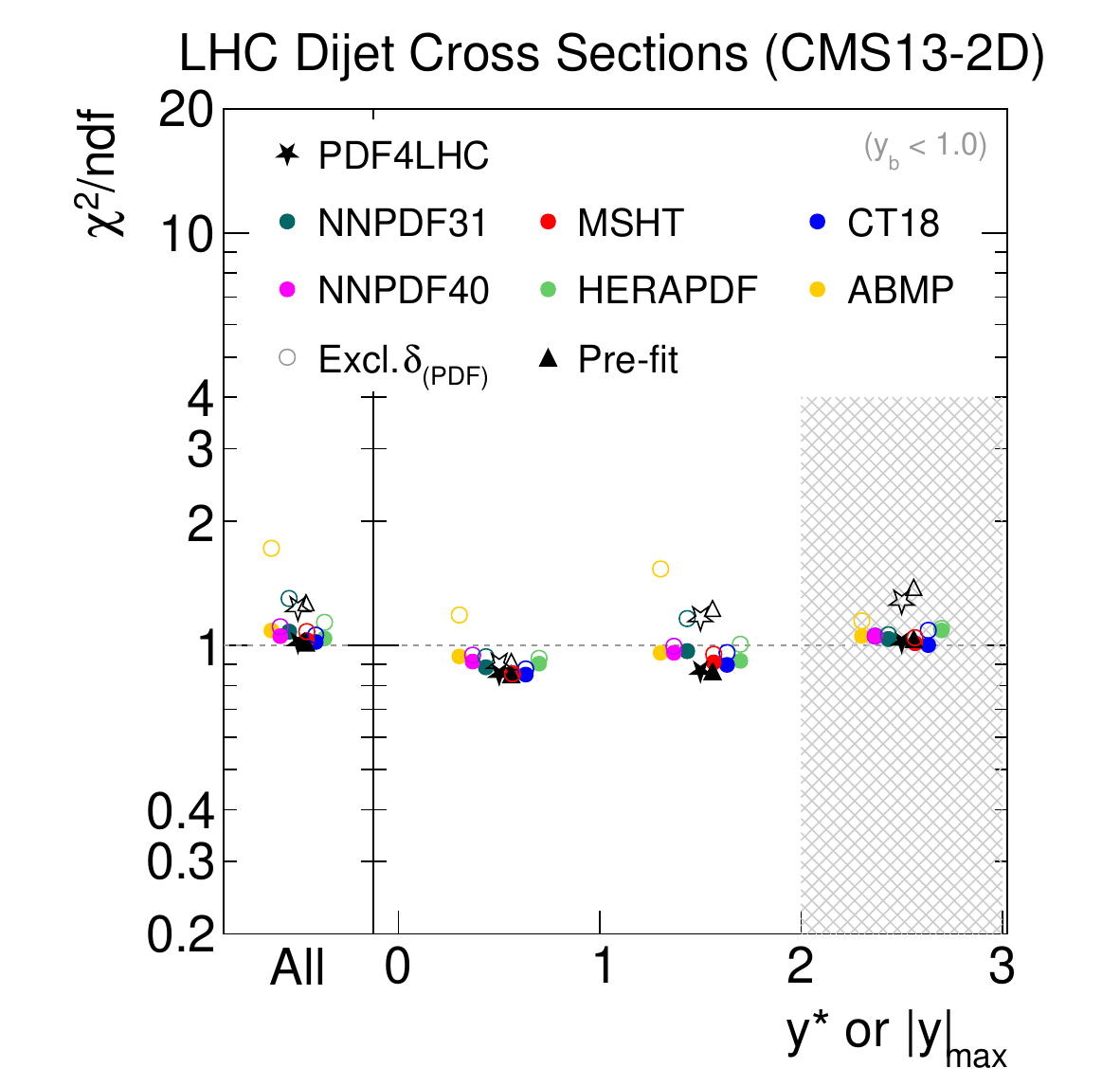} % \linewidth
  \includegraphics[width=0.24\columnwidth,trim={20 0 20 0},clip]{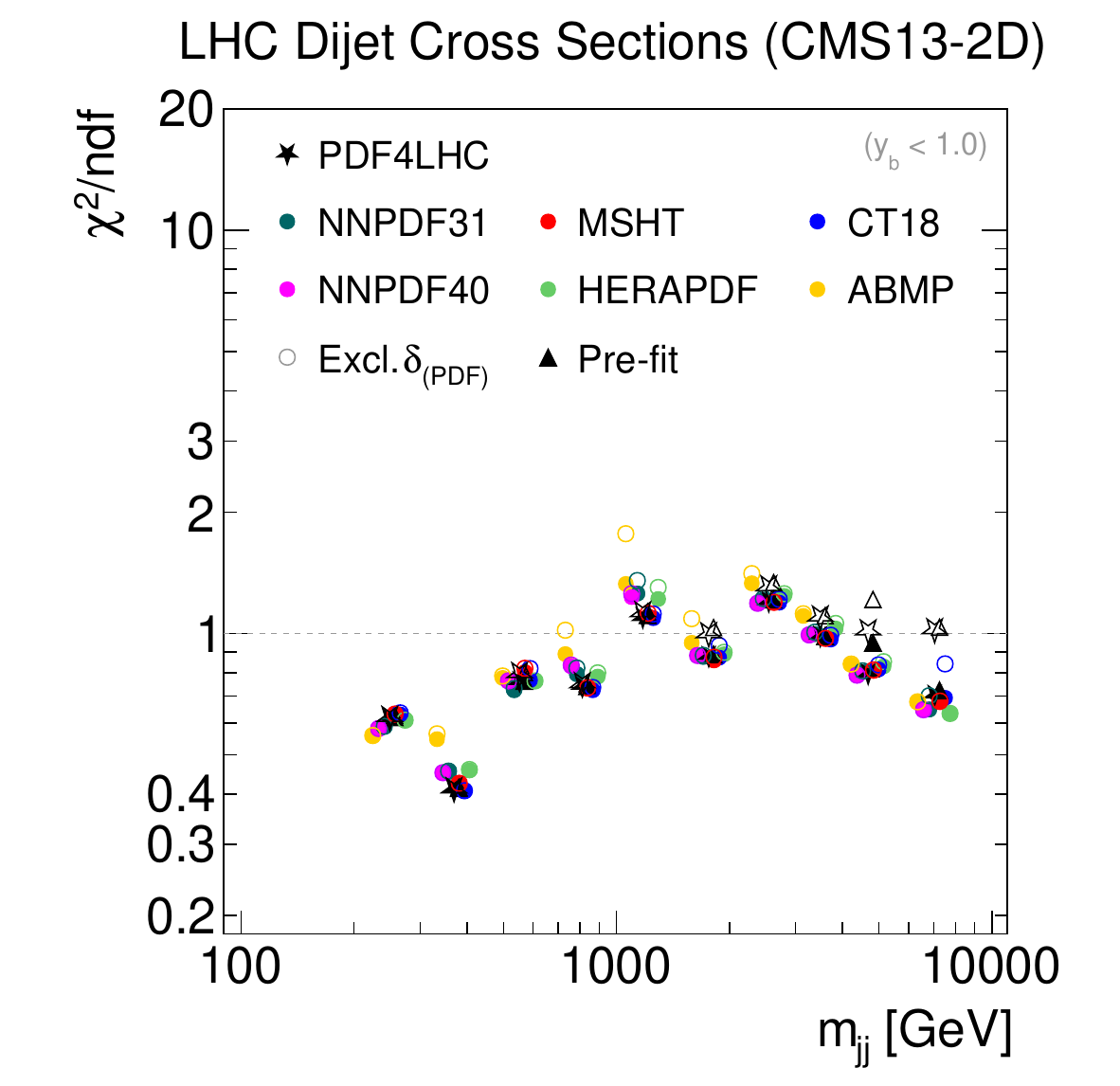} \\ % \linewidth
  \includegraphics[width=0.24\columnwidth,trim={20 0 20 0},clip]{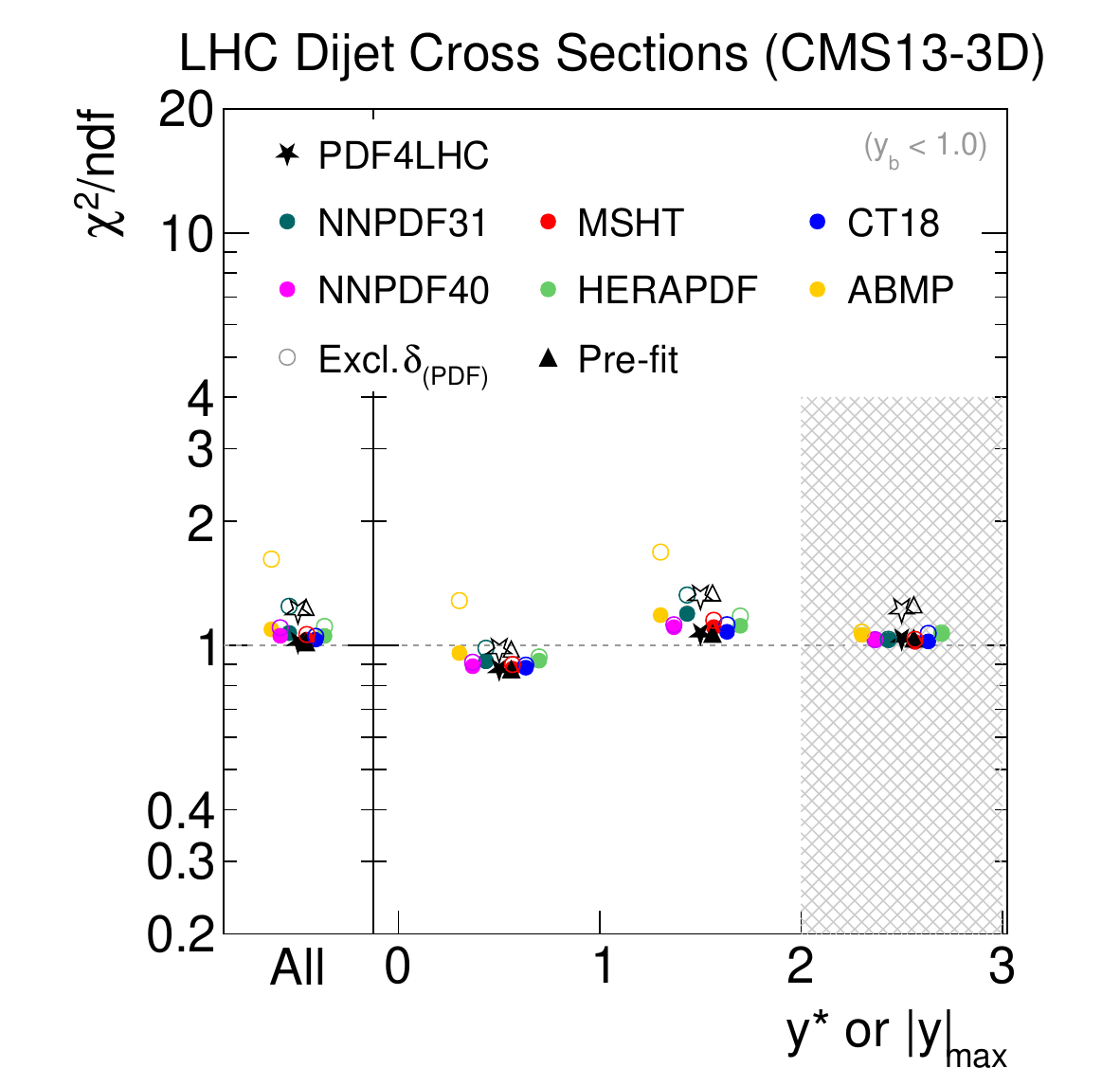} % \linewidth
  \includegraphics[width=0.24\columnwidth,trim={20 0 20 0},clip]{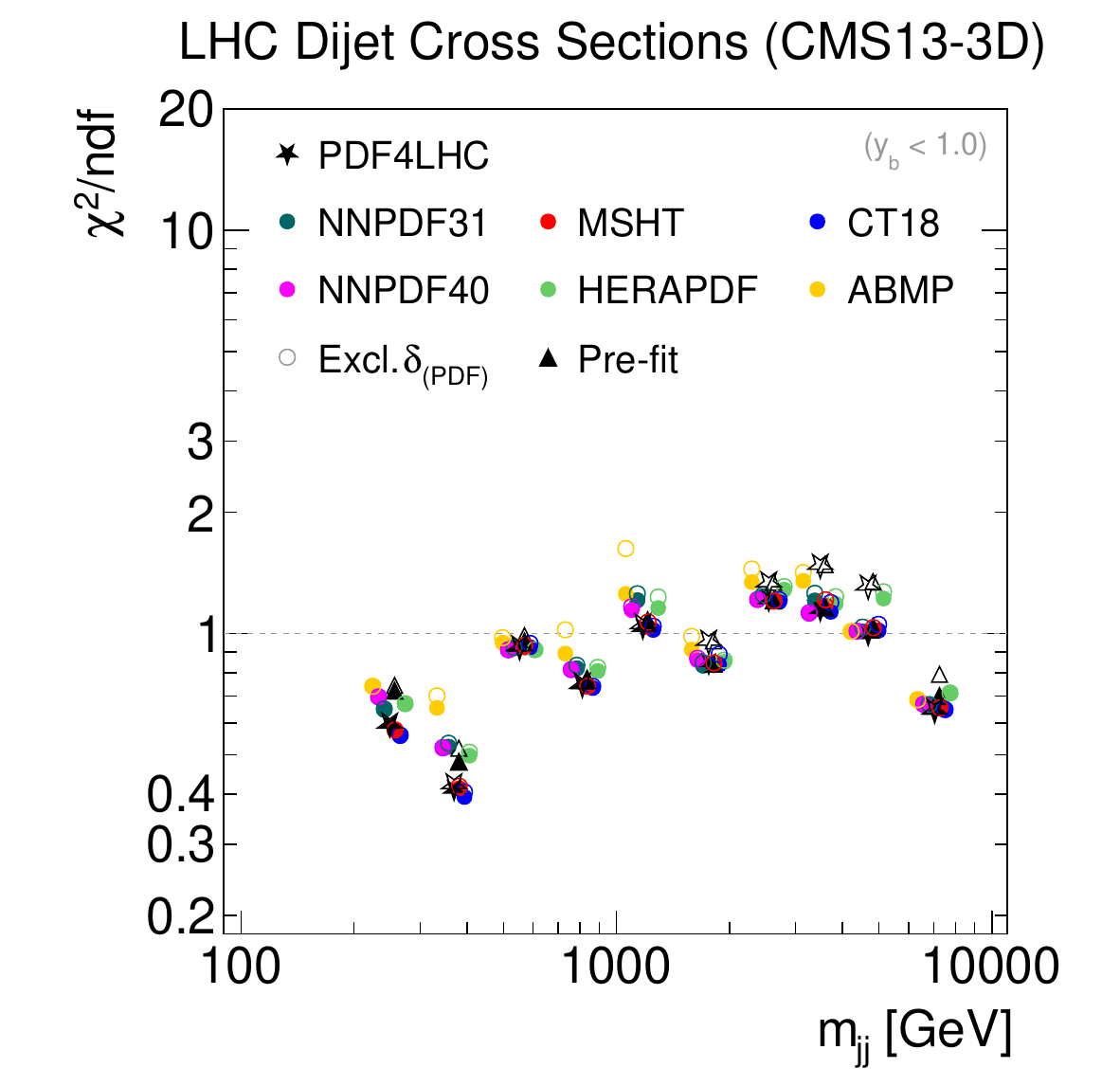} % \linewidth
  \caption{ Left: Post-fit values of \chisq/\ndf of \as-fits in three distinct \ys-ranges ($0\leq\ys<1$, $1\leq\ys<2$, $\ys\geq2$).  More details, see
    Fig.~\ref{fig:chisq}.  Right: Post-fit values of \chisq/\ndf of \as-fits in nine distinct \mjj-ranges.  Excellent consistency of the four data
    sets, and the data and NNLO predictions is observed.  The leftmost entries show the \chisq/\ndf values of the nominal combined fit to all dijet
    data.  }
  \label{fig:chisq-ys}
\end{figure}

The resulting \chisq/\ndf values are displayed in Fig.~\ref{fig:chisq-ys} (left) and excellent \chisq/\ndf values around unity are obtained for all
three \ys ranges and for different PDF sets.
It is also observed that including PDF uncertainties in \chisq alters the \chisq/\ndf values only slightly, which indicates an excellent agreement of the PDFs with the data, as well as small PDF uncertainties.
However, a fit to all three \ys ranges at a time yields a somewhat increased \chisq/\ndf value and thus indicating a slight tension between all data.
In order to avoid a possible bias from that, and to reduce further the PDF dependence, we drop the data with $\ys>2$ (or $\ym>2$, respectively) in the nominal fit.
This restriction removes jets in the outer rapidity regions, where the endcap calorimeters are important and tracking detectors are not available.

In order to assess the consistency of the data across different \mjj regions, nine adjacent ranges between 200\GeV and 9\TeV are defined in \mjj, with an approximately equidistant width in $\log(\mjj)$, similar to the data intervals.
% Each cross section value is identified with a single average value \mjj, which is the logarithmic average of its \mjj bin edges.
For the data of Ref.~\cite{CMS:2017jfq}, which are measured as a function of $p_\text{T,avg}$, the \mjj-interval is sampled with the NNLO calculation, and the
average \mjj values are found to range from 218 to 5396\GeV.  Nine fits to the individual \mjj ranges are performed and the resulting \chisq/\ndf
values are displayed in FIG.~\ref{fig:chisq-ys} (right).  Altogether, reasonable values of \chisq/\ndf are obtained. At lower values of \mjj, the values are below unity, whereas at $\mjj\approx2.5\TeV$ they are somewhat larger with values of about 1.2.  The inclusion of the PDF
uncertainties in the \chisq has only a limited impact on the resulting \chisq/\ndf values, indicating little sensitivity to the PDF parameters
and good agreement with PDFs, given the imposed cuts on \ys and \yb ($ys<2$ and $\yb<1$).

\subsubsection{HERA dijet data}
\begin{figure}[tbh!]
  \includegraphics[width=0.29\columnwidth,trim={20 0 20 0},clip]{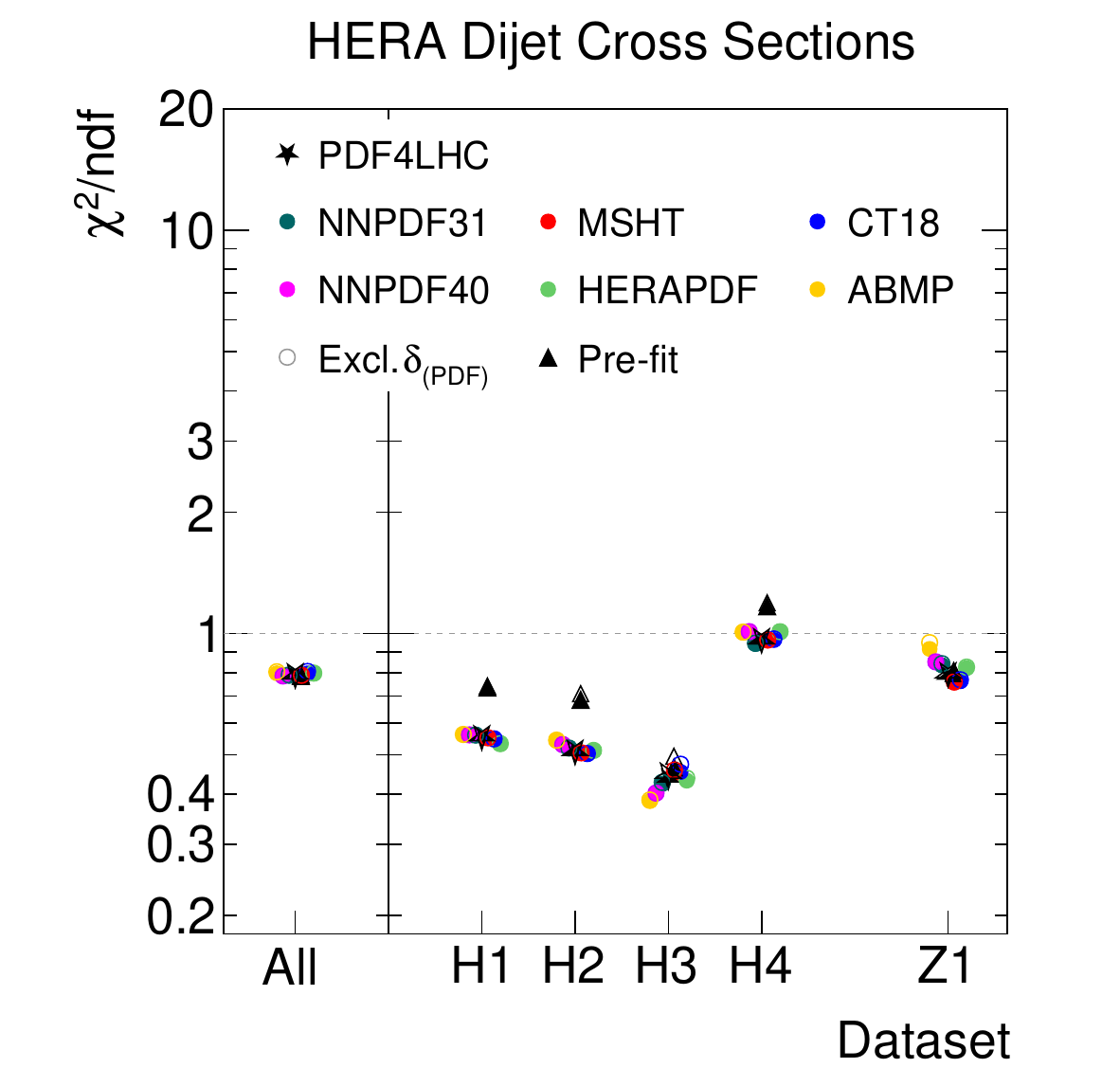} % \linewidth
  \includegraphics[width=0.29\columnwidth,trim={20 0 20 0},clip]{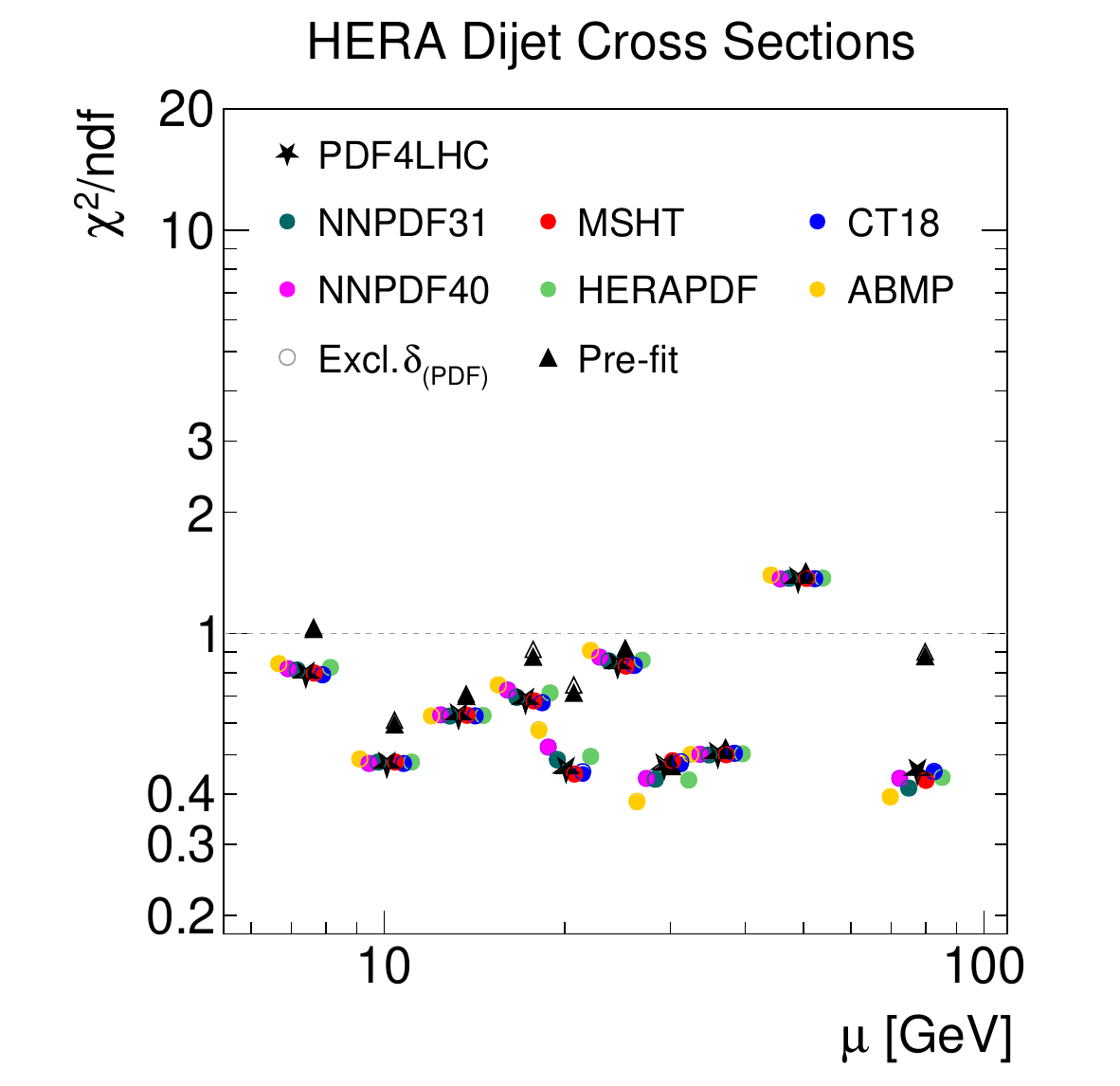} % \linewidth
  \caption{
    Left: Post-fit values of \chisq/\ndf of \as fits for five dijet data sets from HERA.
    Four data sets from H1 for center-of-mass energies and \Qsq\ ranges are studied, and labeled as H1~\cite{H1:2000bqr}, H2~\cite{H1:2010mgp}, H3~\cite{H1:2016goa} and H4~\cite{H1:2014cbm}, and the data set from ZEUS is labeled Z1~\cite{ZEUS:2010vyw}.
    The combined fit to all HERA data is labelled as `All'.
    Right: Post-fit values of \chisq/\ndf of \as-fits to all five HERA dijet data sets in ten distinct $\mu$ ranges.
    See FIG.~\ref{fig:chisq} for more details.
    }
  \label{fig:chisq-hera}
\end{figure}

In this section, we present a study of the consistency of the dijet data from H1 and ZEUS, using the NNLO pQCD predictions.
Similar studies have previosuly been performed by the H1 Collaboration for the H1 data sets~\cite{H1:2017bml}, and for a combination of ZEUS dijet data and selected H1 data sets in Ref.~\cite{H1:2021xxi}.
Nonetheless, we perform a study similar to those performed for the LHC data, considering all H1 dijet data sets~\cite{H1:2000bqr,H1:2010mgp,H1:2016goa,H1:2014cbm} along with the dijet data from ZEUS~\cite{ZEUS:2010vyw}. The \chisq/\ndf value for each data set and multiple PDF sets are displayed in FIG.~\ref{fig:chisq-hera}.

The \chisq/\ndf values for the H1 data sets are very similar to those reported in Ref.~\cite{H1:2017bml}.
Similarly, the \chisq/\ndf value for the ZEUS data confirms the good agreement between the data and the NNLO predictions, as previously reported~\cite{ZEUS:2010vyw}.
The combined fit to all HERA data results in an excellent \chisq/\ndf with a value of 0.79, thus confirming excellent consistency between the different data sets and of the data with the NNLO predictions.
Different PDF sets have only little impact on the \chisq/\ndf values, which may be explained by the strong impact of the HERA inclusive DIS data on PDFs.
Subsequently, the data are grouped into ten $\mu$ intervals with $\mu=\Qsq+\ptavg^2$~\cite{H1:2017bml}.
The resulting \chisq/\ndf values for these fits are also very good.
Although some moderate fluctuations in \chisq/\ndf\ are observed across different $\mu$ intervals, no systematic deterioration is evident.

This study confirms that the HERA dijet data can be used for an unbiased determination of the running of \as together with the NNLO predictions
across their full range.
However, we exclude the lowest $\mu$ interval because it falls below twice the mass of the bottom quark. Our calculations are performed for five massless quark flavors and are therefore not strictly valid for these data, although they still provide an excellent description of them.

%\end{widetext}

\end{document}